\documentclass[12pt]{article}

\usepackage[T1]{fontenc}
\usepackage[utf8x]{inputenc}
\usepackage[english]{babel}
\usepackage{cite}
\usepackage{hyperref}
\usepackage{physics}
\usepackage{epstopdf}
\usepackage{fancyhdr} 
\usepackage{pict2e}
\usepackage{tabularx}
\usepackage[percent]{overpic}
\usepackage{wrapfig}
\usepackage[absolute]{textpos}
\usepackage{graphicx}
\usepackage{epstopdf}
\usepackage{caption}
\usepackage{subcaption}
\usepackage{adjustbox}
\usepackage{xcolor} 
\usepackage{enumerate} 
\usepackage{geometry} 
\usepackage{amsmath} 
\usepackage{amssymb}
\usepackage{IEEEtrantools}

\title{Generalizing the Method of Images for Complex Boundary Conditions : Application on the LHC Beam Screen}
\author{P. Bélanger \footnotesize{*(Grad. Research Assistant, Beam Physics group at TRIUMF and TE-MPE-PE at CERN)}}
\date{\today}

\begin{document}
\bstctlcite{IEEEexample:BSTcontrol}

	\pagenumbering{gobble}
	\makeatletter	
	\begin{textblock}{10}(4.1,1)
		\begin{flushright}
			\today \\
			\url{philippe.belanger@cern.ch}\\
			\url{pbelanger@triumf.ca}
		\end{flushright}
	\end{textblock}
	
	\phantom{e}\vspace{20mm}
	{\Large{\textbf{\@title}}}\vspace{1mm}\\
	{\@author} \\
	
	
	\vspace{-5mm}
	\noindent\rule{\textwidth}{0.5pt}\vspace{-4.5mm}
	\noindent\rule{\textwidth}{0.5pt}
	
	\vspace{-5mm}
	\subsection*{Summary}
	This paper seeks to show that the beam screen of the LHC has an important effect on the electric field of the LHC beam, a few tens of sigmas away from its center. To do so, we develop two new methods for finding the effect of a complex conducting boundary for boundary value problems in electrostatics. Both methods are based on a generalization of the method of images and require low computing power. The result is an exact solution to the problem of a discretized conducting boundary, which we take to be an approximation of the real solution. As an application, we compute the total electric field inside the LHC beam screen and show that neglecting the effect of the conducting boundary is only accurate to 1\% for locations closer than $10\sigma$ from the center of the beam, and only accurate to 10\% for locations closer than $30\sigma$.

	
	\noindent\rule{\textwidth}{0.5pt}
	
	\tableofcontents
	\vspace{-50mm}

	\newpage
\pagenumbering{arabic}
\setcounter{page}{2}

\section{Context}
This paper seeks to show that the beam screen of the LHC has an important effect on the electric field of the LHC beam, a few tens of sigmas away from its center. To do so, we develop a generalization of the method of images to solve boundary value problems in electrostatics and compute the difference between the free space electric field coming from a 2D gaussian beam and the actual total field in presence of the LHC beam screen.\\

The LHC beam is surrounded first by a beam screen and then by a beam pipe, as shown on figure \ref{fig:beamScreen}. The presence of this beam screen makes the problem of finding the electric field a \textit{boundary value} problem, since the potential must be constant over the whole screen.

\noindent
\begin{minipage}{.6\textwidth}
  	\captionsetup{width=0.9\textwidth}
	\centering
	\includegraphics[width = 0.7\linewidth]{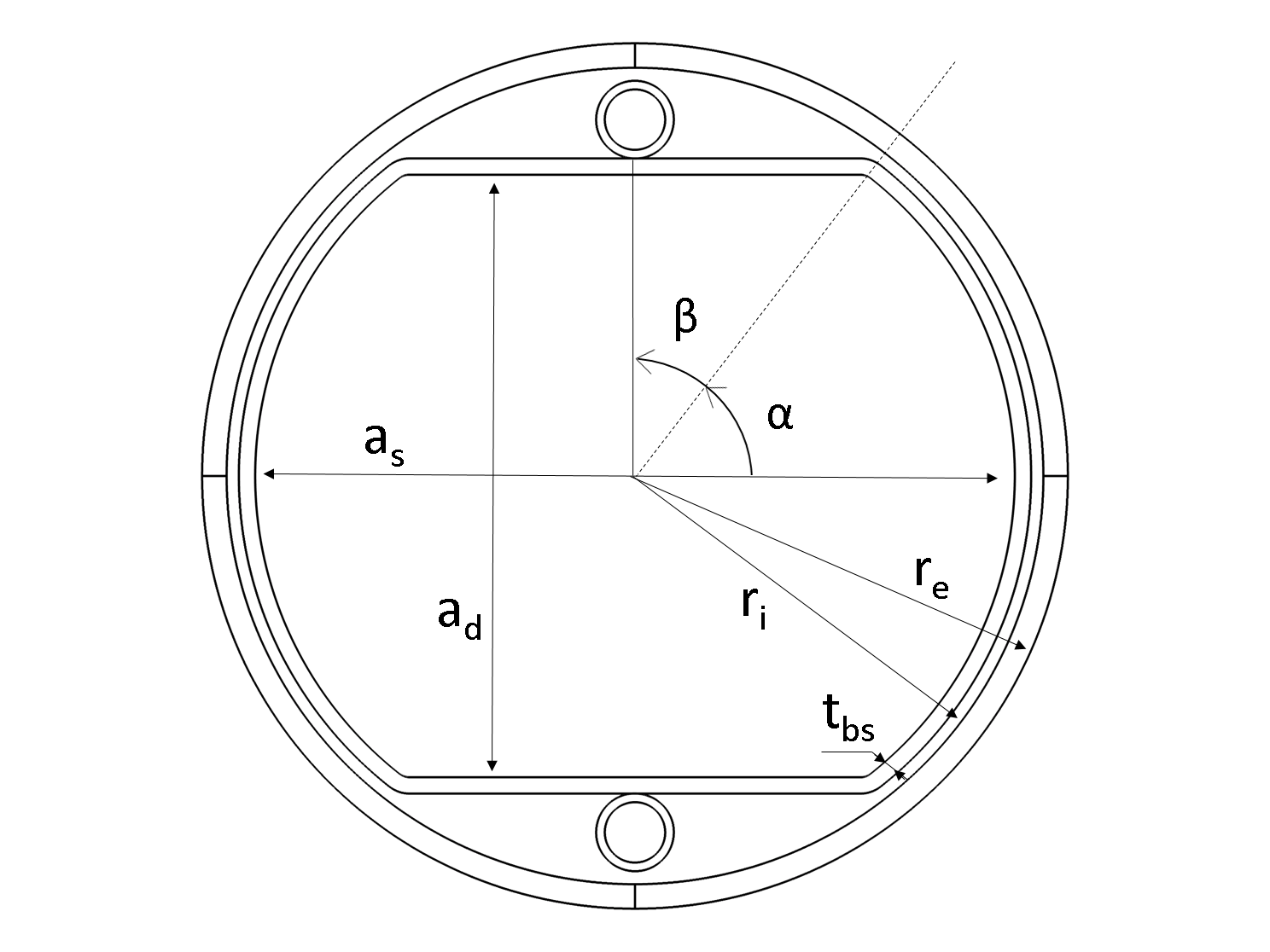}
	\captionof{figure}{LHC Beam screen cross section \cite{Morrone2017}}
	\label{fig:beamScreen}
\end{minipage}
\begin{minipage}{.4\textwidth}
\centering
\captionof{table}{Beam screen dimensions}
  \begin{tabular}{clcccc}
			\hline
			Symbol & Value\\\hline\hline
			$a_d$ & 36.9 mm\\\hline
			$a_s$ & 46.5 mm\\\hline
			$\alpha$ & 52.4$^\circ$\\\hline
			$\beta$ & 37.6$^\circ$\\\hline
			$r_i$ & 25.0 mm\\\hline
			$r_e$ & 26.5 mm\\\hline
			$t_{bs}$ & 1.00 mm\\\hline
		\end{tabular}
\end{minipage}

\section{Free space E-field of a 2D gaussian charge distribution}
As stated by B. Auchmann in \cite{Auchmann}, the Lorentz factor in the charge density of a relativistic continuous proton beam  is balanced by the same factor in the Lorentz transformation of the field. The transverse electric field can therefore be computed as if the proton charge distribution was a rest, uniformly distributed over the LHC ring. Also (from \cite{Koscielniak2000}), in the 2D electrostatic representation of our problem, we must distinguish the contributions of charged filaments of infinite length and linear charge density $\lambda$, for which (from Gauss' law) :
\begin{equation}
E_r = \frac{\lambda}{2\pi\varepsilon_0 r}\qquad\text{and}\qquad \Phi  =-\frac{\lambda}{2\pi\varepsilon_0}\ln(r)
\end{equation}

from the contributions of point charges $Q$ for which :
\begin{equation}
E_r = \frac{Q}{4\pi\varepsilon_0 r^2}\qquad\text{and}\qquad \Phi  =\frac{Q}{4\pi\varepsilon_0 r}
\end{equation}

Obviously, the proton beam of the LHC can be modelled, in a first approximation, by a infinite charged filament with $\lambda = N_p/C$. As a second approximation, one can use the Houssais potential\footnote{Commonly known as the Bassetti-Erskine formula. The original expression is, however, coming from D. Houssais as recently discussed in \cite{Koscielniak2019}. Bassetti and Erskine mostly worked on a numerical calculation of the field.} to find a really accurate description of the electric field coming from an asymmetric 2D gaussian charge distribution (see \cite{Auchmann}). Below is a comparison of the free space field from the Houssais E-field for an asymmetric beam, a round beam and the field from an infinite charged filament.

\begin{figure*}[!h]
	\captionsetup{width=0.9\textwidth}
	\centering
	\begin{subfigure}[t]{0.32\linewidth}
		\centering
		\includegraphics[width = 1.1\linewidth]{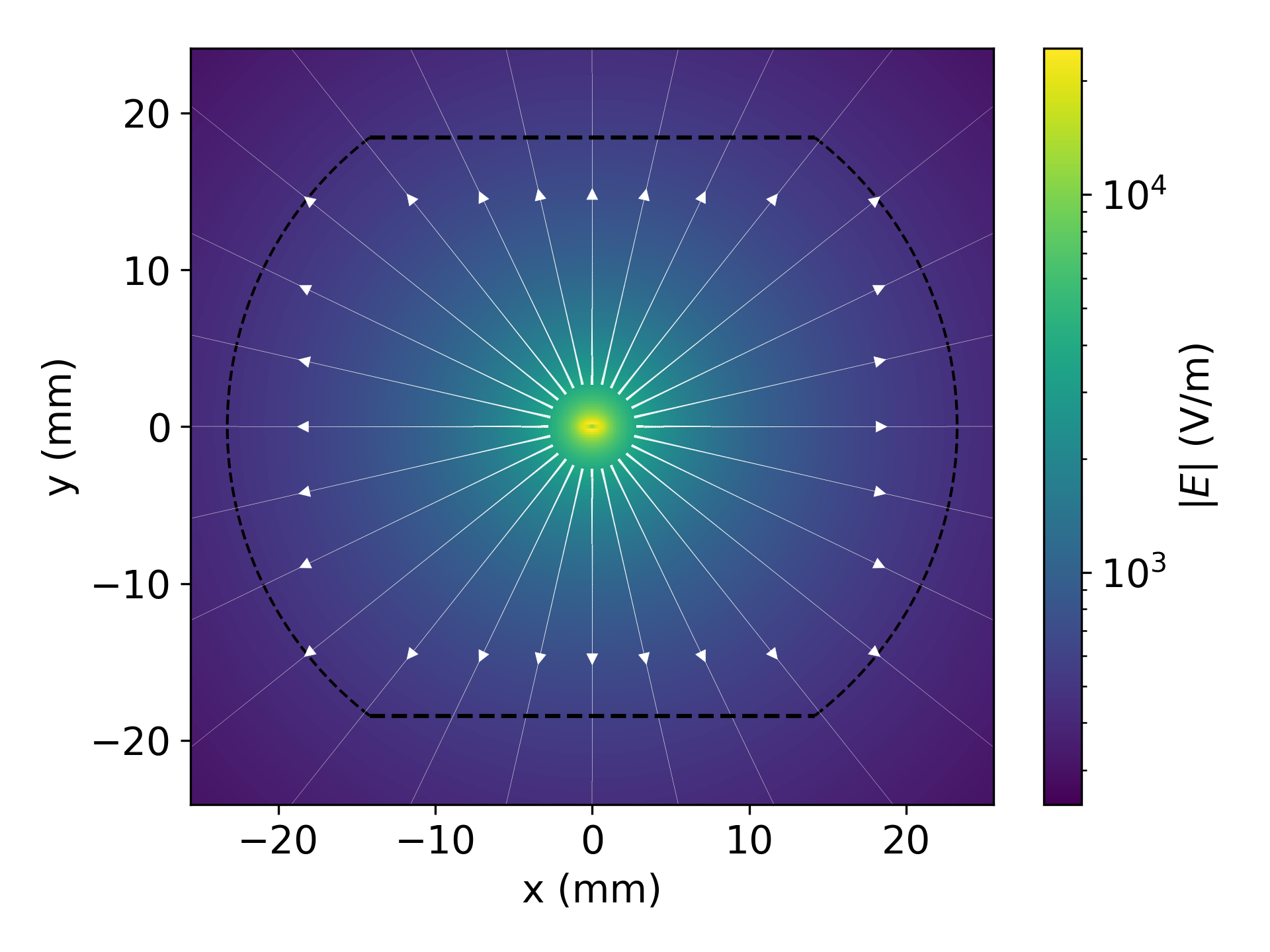}
		\caption{}
	\end{subfigure}%
	~\hspace{0mm}
	\begin{subfigure}[t]{0.32\linewidth}
		\centering
		\includegraphics[width = 1.1\linewidth]{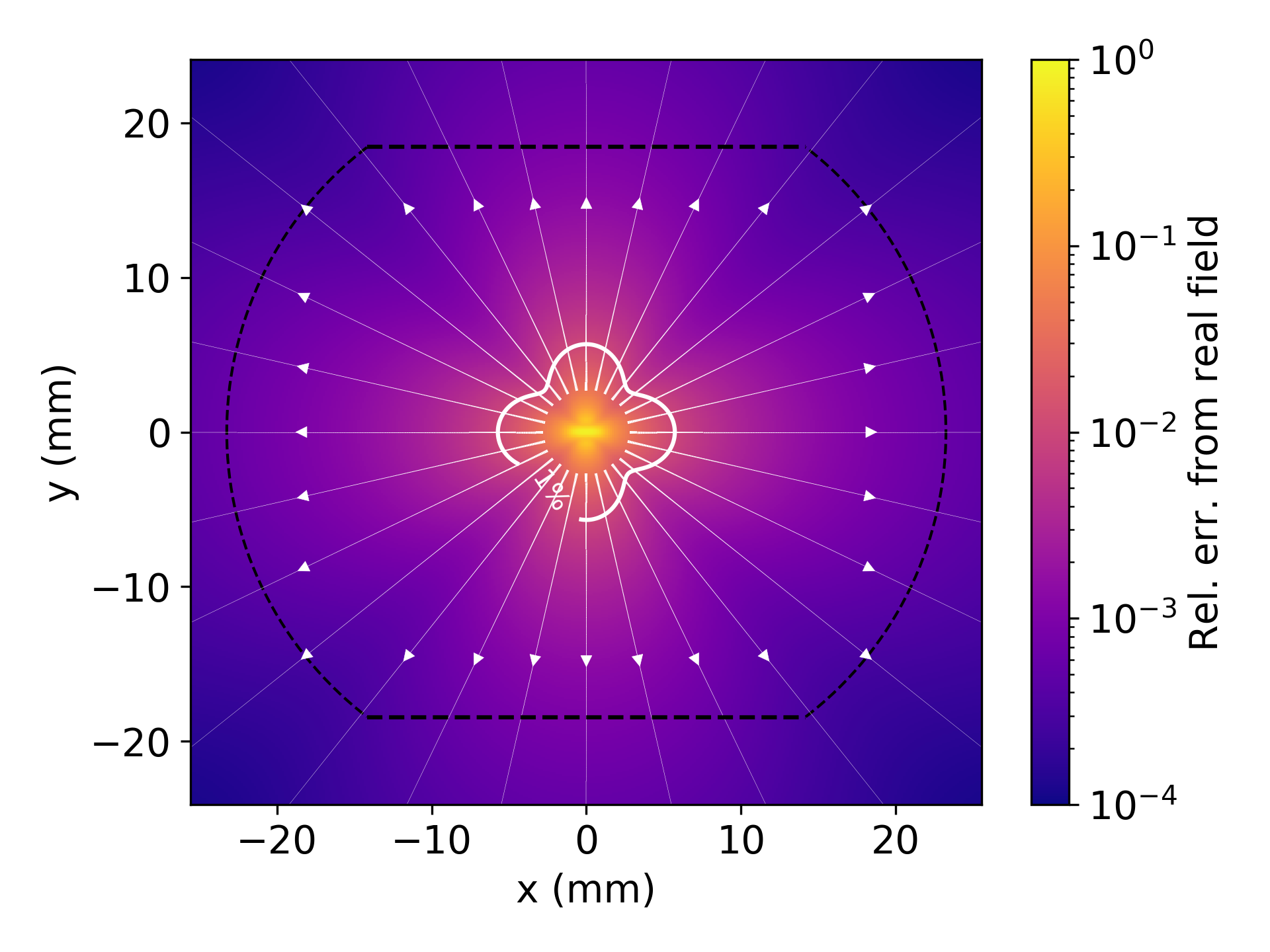}
		\caption{}
	\end{subfigure}
	~\hspace{0mm}
	\begin{subfigure}[t]{0.32\linewidth}
		\centering
		\includegraphics[width = 1.1\linewidth]{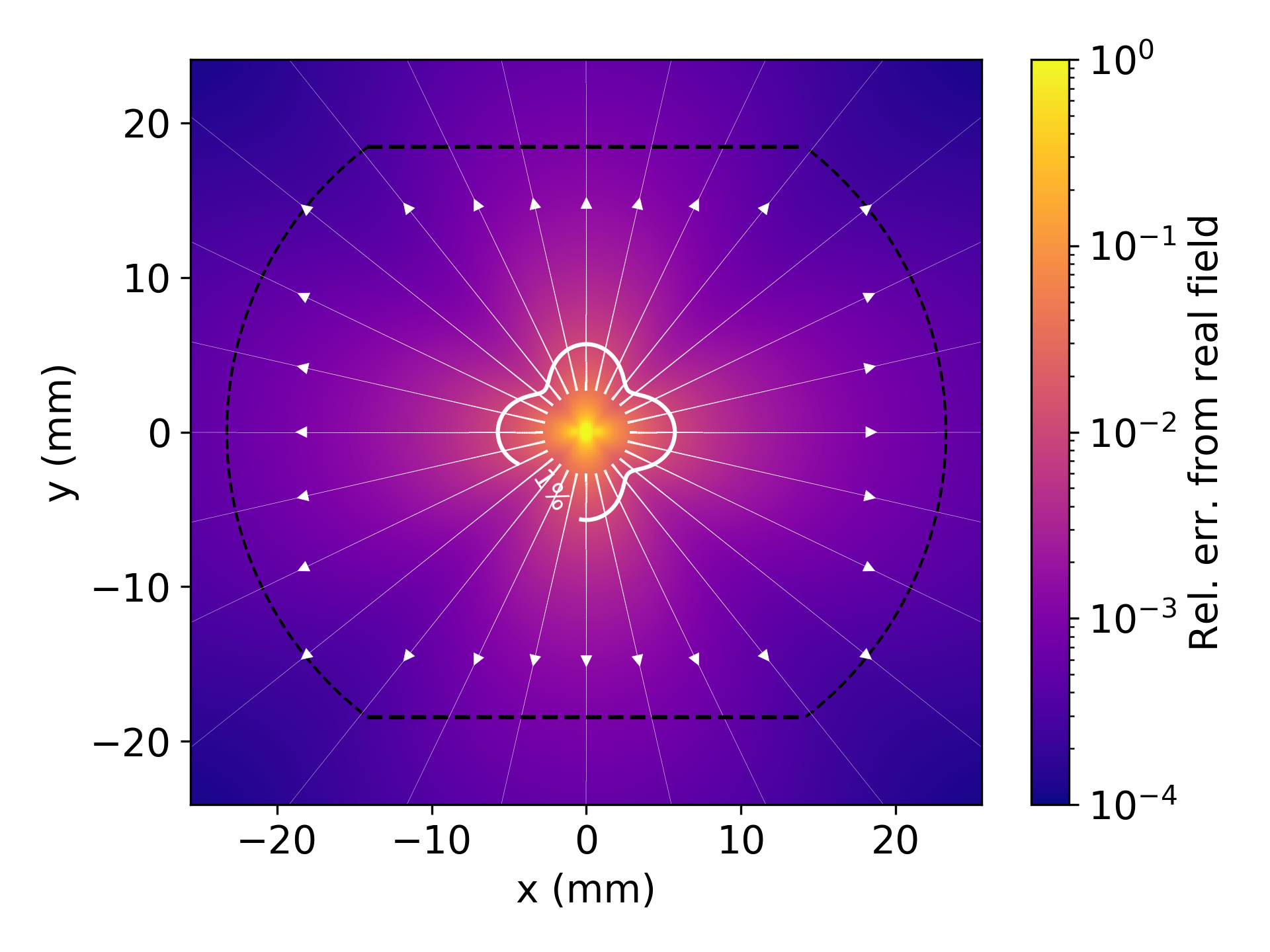}
		\caption{}
	\end{subfigure}
	\caption{\small{Comparison of (a) The free space Houssais E-field of a beam with $\sigma_x = 335\ \mu$m, $\sigma_y = 105\ \mu$m; (b) The free space Houssais E-field of a round beam with $\sigma_x=\sigma_y = 335\ \mu$m; (c) The free space E-field of an infinite charged filament. Plots (b) and (c) show the magnitude of the relative error vector (component by component) on the color bar $Err = |(\vec E_{beam} - \vec E_{approx})/\vec E_{beam}|$. The level curve for an error of 1\% is shown. Outside this region, the difference between all fields is deemed negligible.}}
	\label{fig:fieldComparison}
\end{figure*}
	
In the previous figure, the beam screen is represented for reference only and is not considered in the calculations. However, \textbf{we clearly see that the free space fields are incompatible with our boundary value problem, as the field lines (shown in white) are not perpendicular to the surface of the beam screen}. The following sections try to address this problem.


\section{Method of images for solving boundary value problems}
\subsection{Quick overview}

\textbf{The main interest for this method is coming from its simplicity and its light computing requirements}. From Wikipedia : The method of images  (or method of mirror images) is a mathematical tool for solving differential equations, in which the domain of the sought function is extended by the addition of its mirror image with respect to a symmetry hyperplane. As a result, certain boundary conditions are satisfied automatically by the presence of a mirror image, greatly facilitating the solution of the original problem.\\

In the case of electrostatics, Maxwell's equations must always be respected. For our problem, Gauss' law and Poisson's equation are particularly important:
\begin{equation}
\oint_S \vec E \cdot d\vec A = \frac{Q_{int}}{\varepsilon_0}\qquad \text{and}\qquad \nabla^2 \Phi = -\rho/\varepsilon_0
\end{equation}

When a conductor like the beam screen is placed around a charge, the charge distribution in the conductor rearrange in such a way that all field lines meet the conductor orthogonally and that the potential is uniform on the whole surface of the conductor. With the method of images, we forget about this conductor and place image charges outside of our \textit{physical boundary} (automatically respecting Gauss' law), hoping to mimic the effect of the conductor on the total field inside the \textit{physical boundary}. In 3D, adding any number of point charges doesn't change the Laplacian inside the boundary. In 2D, the same can be said about infinitely charged filaments. It can be shown that the solution to Poisson's equation is unique. Therefore, if we find a distribution of image charges that allows us to respect the boundary condition, we have \textit{the} solution to Poisson's equation and the problem is solved.  

\subsection{Known case 1 : Infinite parallel conducting planes}

In the case of two infinite, parallel, conducting planes, the method of images requires an infinite number of image charges since each charge added by reflection on a given plane produces another image charge by reflection on the second plane, and so on. If the original real charge is located at the center of the two planes ($y = L/2$), then the solution is to put image charges $Q'$ like so :
\begin{equation}
\left\{\begin{aligned}
&Q' = Q&\qquad&\text{at}\qquad y' = 2nL + y\qquad\text{for}\ n\in \mathbb{Z}^*\\
&Q' = -Q&\qquad&\text{at}\qquad y' = 2nL - y\qquad\text{for}\ n\in \mathbb{Z}
\end{aligned}\right.
\end{equation}

The same can be said about an infinite charged filament  located at the center of the plates by replacing $Q$ with $\lambda$. If we set our origin on the center of such a filament and place the planes at $y = \pm L/2$, we find that the total potential is (including the contribution from the real charge):
\begin{equation}
\begin{aligned}
\Phi(\vec r) &= -\frac{\lambda}{2\pi\varepsilon_0}\left[\sum_{-\infty}^{\infty}\ln\Big(|\vec r - 2nL\hat y|\Big) - \ln\Big(|\vec r - L(2n-1) \hat y|\Big)\right]\\
&= -\frac{\lambda}{2\pi\varepsilon_0}\left[\ln\Big(\vec r\Big) - \ln\Big(\vec r + L\hat y\Big) - \ln\Big(\vec r - L\hat y\Big) + \dots\right]
\end{aligned}
\label{eq:potentialPlane}
\end{equation}


\subsection{Known case 2 : Infinite conducting cylinder}

In the case of an infinite charged filament surrounded by a cylindrical conductor, only one image filament is required to respect Poisson's equation. If the original real filament is located at the center, the image is located at infinity (i.e. the boundary condition is already satisfied). If the original real filament is at a distance $\vec r$ from the center of  the cylinder (of radius $a$) the problem can be nicely solved using the circle of Apollonius (from the Greek mathematician, Apollonius of Perga). It can be shown that the image filament must be:
\begin{equation}
\lambda' = -\lambda \qquad \text{at}\qquad \vec{r'} = \frac{a^2}{|\vec r|^2}\vec r
\end{equation}

\subsection{Known case 3 : Conducting spherical shell}
The conducting spherical shell is a third case easily solved with the image charge method and found in any textbook. It shares some particularities with the infinite conducting cylinder but is not really useful to our case, since our problem lack any spherical symmetry.

%

\newpage
\subsection{The case of the LHC beam screen}
The difficulty with the beam screen of the LHC is the low symmetry of its geometry. At first, it seems possible to decompose it in a superposition of the first two cases : two infinite conducting planes and an infinite conducting cylinder. By doing so, the image charges from the planes will lead to image charges \textit{inside} the cylinder upon reflection on the cylinder's conducting surface. However, as stated above, the method of images only works if we put the image charges outside of our \textit{physical boundary}. Putting images inside the cylinder will change the physics of the problem (from Gauss' law) and can't be allowed.\\

To find the solution to our problem, we have to remind ourselves about the main requirement: we want the beam screen to fall on an equipotential line from the total field. After some thinking, we find that a rectangular boundary conditions of the same height then the beam screen but with a wider width could help us achieve this goal. Indeed, with such a boundary condition, the equipotential lines will slowly evolve from circles (close to the beam) to rectangles (outside of the beam screen). With the proper width, we can hope to find some equipotential that matches the beam screen's geometry.\\

\begin{figure*}[!h]
	\captionsetup{width=0.9\textwidth}
	\centering
	\begin{subfigure}[t]{0.45\linewidth}
		\centering
		\includegraphics[width = 0.75\linewidth]{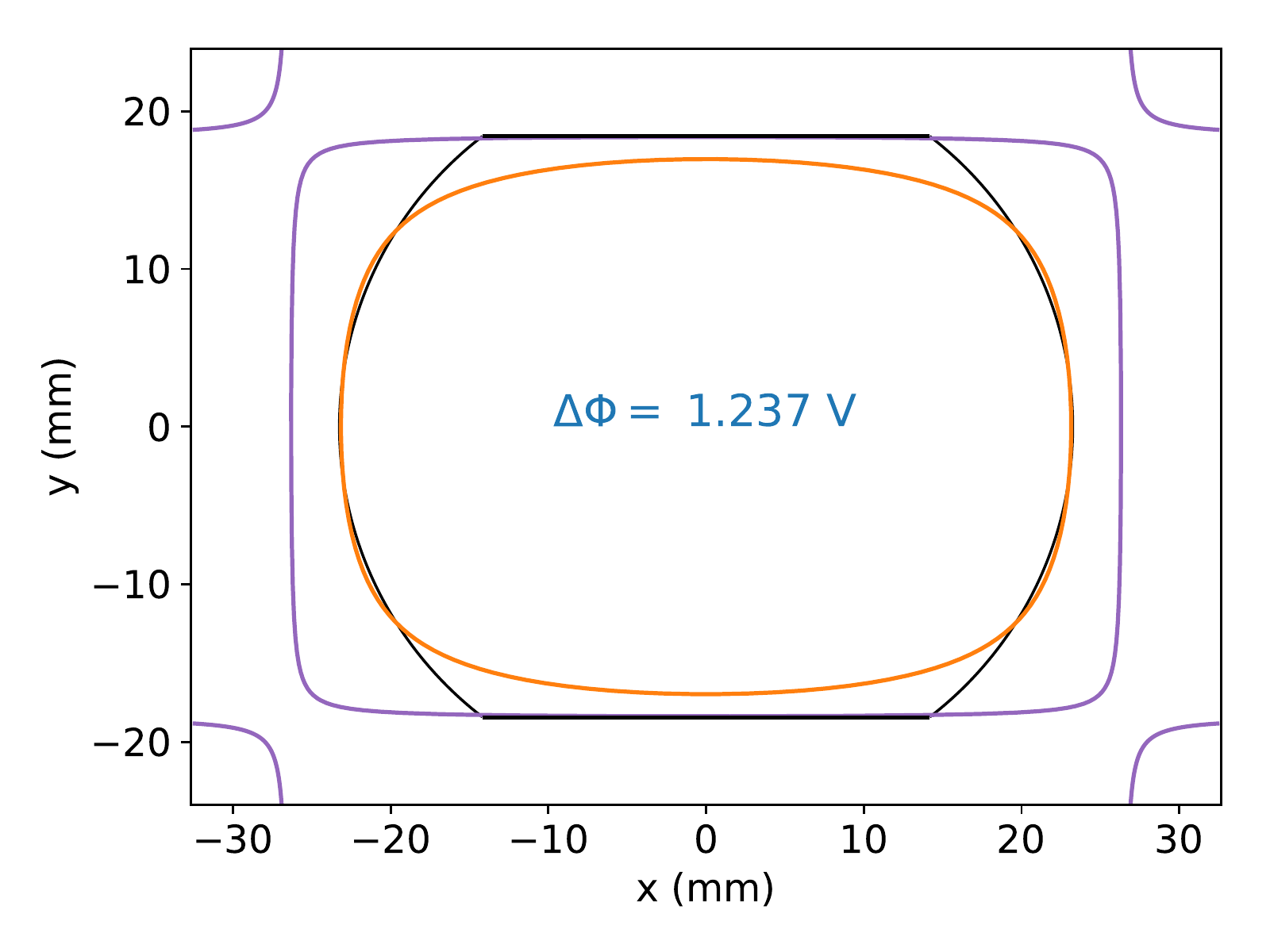}
		\caption{}
		\label{fig:goodPotential}
	\end{subfigure}%
	~\hspace{2mm}
	\begin{subfigure}[t]{0.45\linewidth}
		\centering
		\includegraphics[width = 0.75\linewidth]{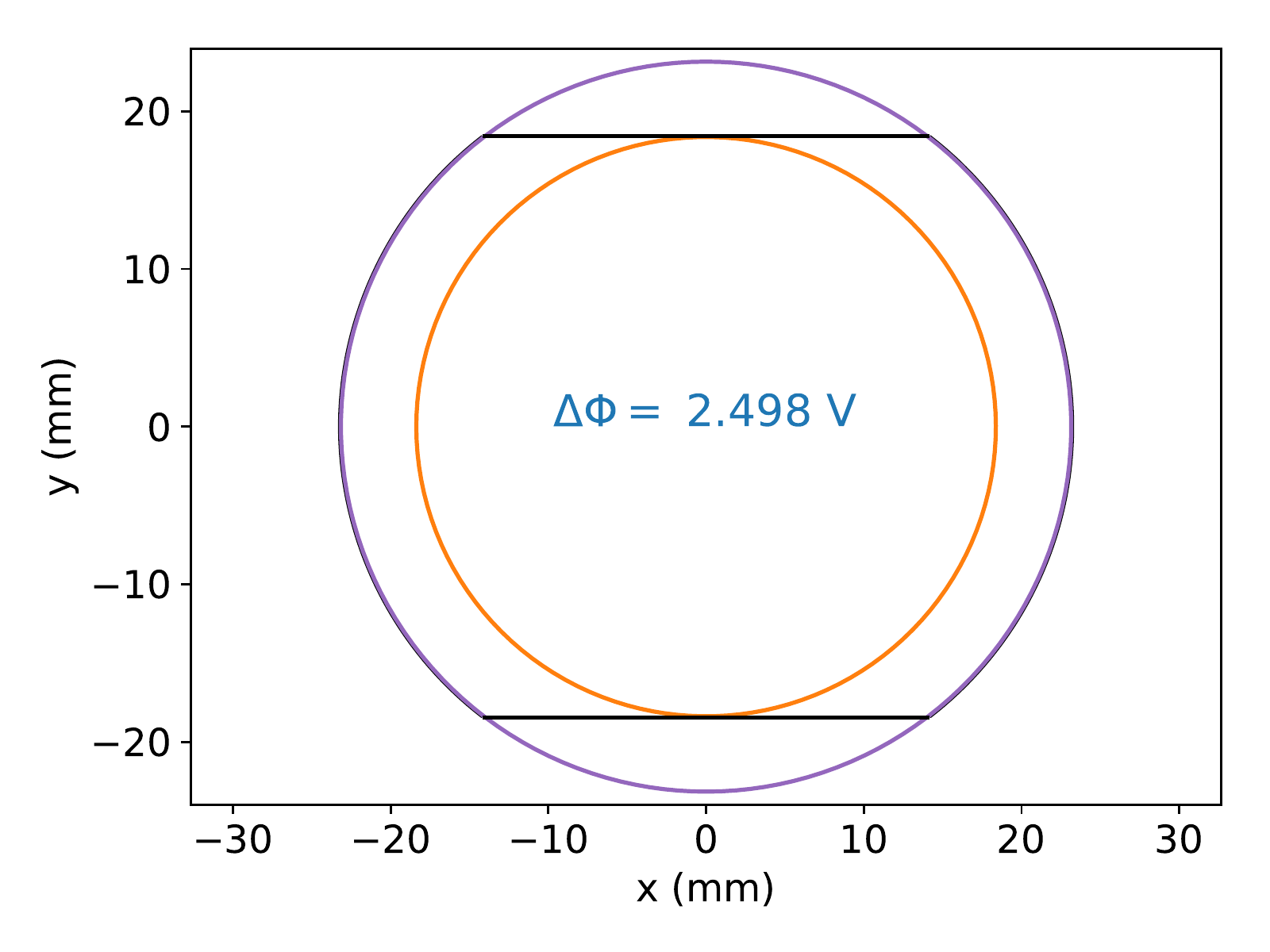}
		\caption{}
		\label{fig:badPotential}
	\end{subfigure}
	\caption{\small{Maximal and minimal equipotential lines touching the beam screen for (a) the beam field with image charges (b) the free space beam field. The difference of potential between each equipotential gives the maximum variation of the potential over the beam screen. This result was obtained with a boundary width of $1.14\cdot a_s$, 20 charges reflected vertically and 2 charges reflected horizontally.}}
	\label{fig:potentials}
\end{figure*}

To study this idea, it is worth working with potentials instead of fields. As shown above, the infinite charged filament (for which we know the analytic expression of the potential) mimics quite accurately the beam field away from the center. For our rectangular boundary condition, all we have to do is apply the case of the infinite parallel conducting plane twice: once vertically, once horizontally. Calculating the potential coming from all the vertical and horizontal reflections of the beam located at the center of the beam screen, we get the result shown on figure \ref{fig:goodPotential}. The conclusion here is obvious: our description of the total electric potential inside the LHC beam screen is more accurate than the free space approximation. This statement is backed up by the fact that 1 - the equipotential lines match the beam screen more accurately and that 2 - the variation of the potential over the whole beam screen is smaller compared with the free space approximation.

\begin{figure*}[!h]
	\captionsetup{width=0.9\textwidth}
	\centering
	\begin{subfigure}[t]{0.45\linewidth}
		\centering
		\includegraphics[width = 1.1\linewidth]{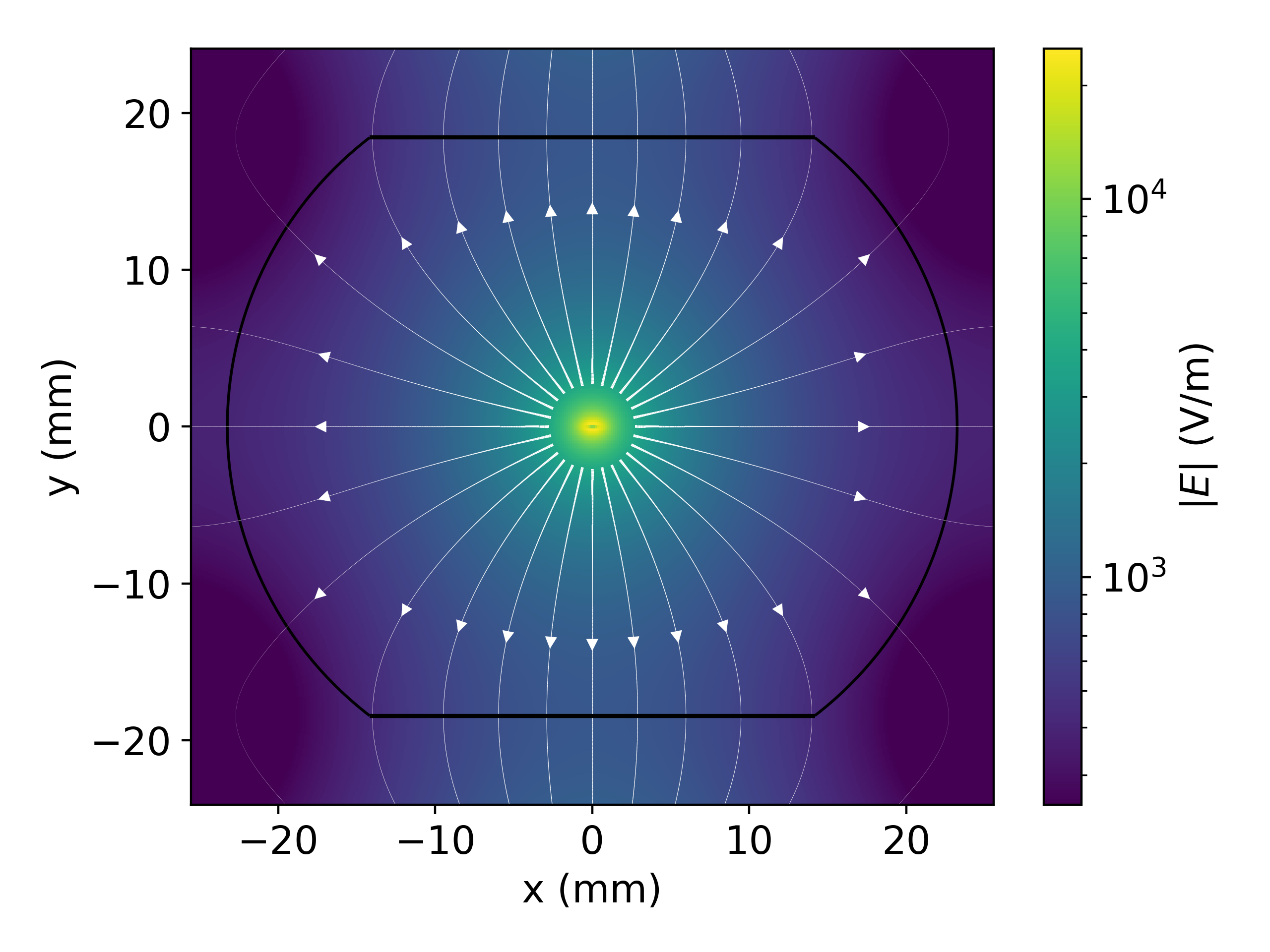}
		\caption{}
		\label{fig:totalField}
	\end{subfigure}%
	~\hspace{7mm}
	\begin{subfigure}[t]{0.45\linewidth}
		\centering
		\includegraphics[width = 1.1\linewidth]{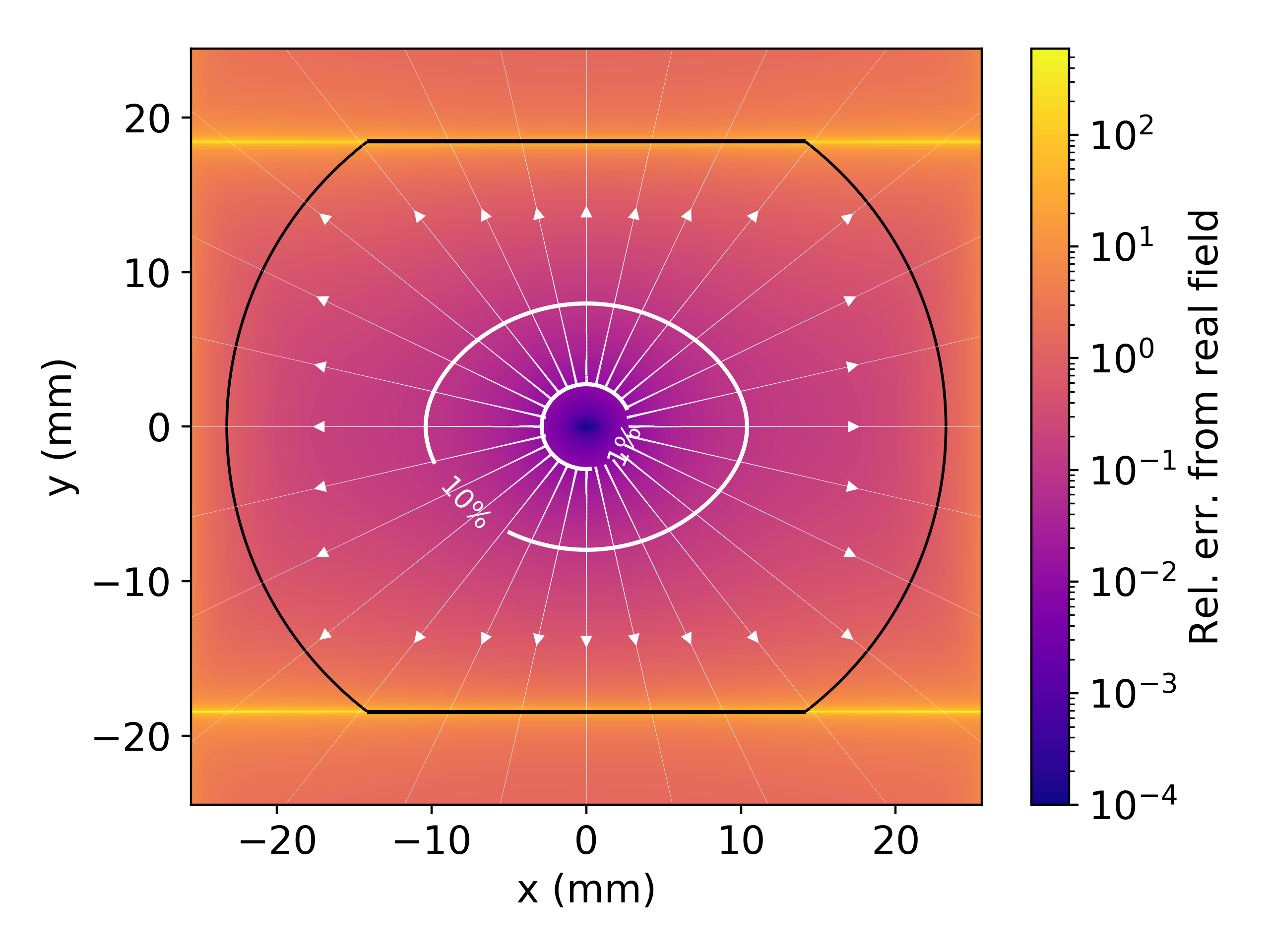}
		\caption{}
		\label{fig:totalError}
	\end{subfigure}
	\caption{\small{Comparison of (a) the total electric field around the beam in presence of the beam screen obtained with the image charge method (b) the free space Houssais E-field. The magnitude of the relative error vector (free space field compared with the total field) is shown on the color bar of (b). The level curves for errors of 1\% and 10\% are shown. Outside those regions, the free space electric field differs greatly from the total electric field and is no longer a good approximation.}}
	\label{fig:ImageField}
\end{figure*}

Following what was shown above, we can now compute the electric field for both cases. The Houssais E-field is used for the contribution of the beam itself, and infinite charged filaments are considered for the images of the beam. The results are shown on figure \ref{fig:ImageField}.

Those results, even if interesting, are not perfect. In the following section, we work on a generalization of the image charge method for complex geometries like the LHC beam screen.

\newpage
\section{Effective surface charge distribution : a novel approach}

Inspired by the image charge method, we can look for a more powerful and more general method to numerically solve boundary value problems in electrostatics. Of course, finite elements could be used, but they lack the computing efficiency of the image charge method, where the overall potential is obtained via a simple summation over multiple charges.\\

As mentioned above, when a charge is surrounded by a conductor, the free charges in the conductor rearrange themselves in such a way that the potential is uniform on the whole surface of the conductor. For really simple geometries, the image charge method gives us a way to mimic the effect of those surface charges and find the solution to our problem. For more complex geometries like the LHC beam screen, the results are, at best, a rough approximation of the real solution. \textbf{In the following sections, we show that placing an \textit{effective continuous charge distribution} just outside of our conducting boundary allows us to find a much more accurate solution without scarifying the computing perks of the image charge method.} Two methods are explored : a relaxation method and a matrix method.

\subsection{Method 1 : Relaxation}
Let $\ell$ be the position along a (2D) conductor evolving alongside the polar angle $\theta$. Placing a real charge somewhere inside of our physical boundary, it is fairly straightforward to evaluate the potential of this charge along the conductor, $\Phi_0(\ell)$ before the equilibrium is reached (i.e. with no other charges at play). Considering the case of a filament $\lambda_Q$ inside a cylindrical boundary of radius $R$ at a position $y=0.5R$, we get the following :

\begin{figure*}[!h]
	\captionsetup{width=0.9\textwidth}
	\centering
	
	\begin{subfigure}[t]{0.45\linewidth}
	\hspace{-10mm}
		\centering
		\includegraphics[width = 0.8\linewidth]{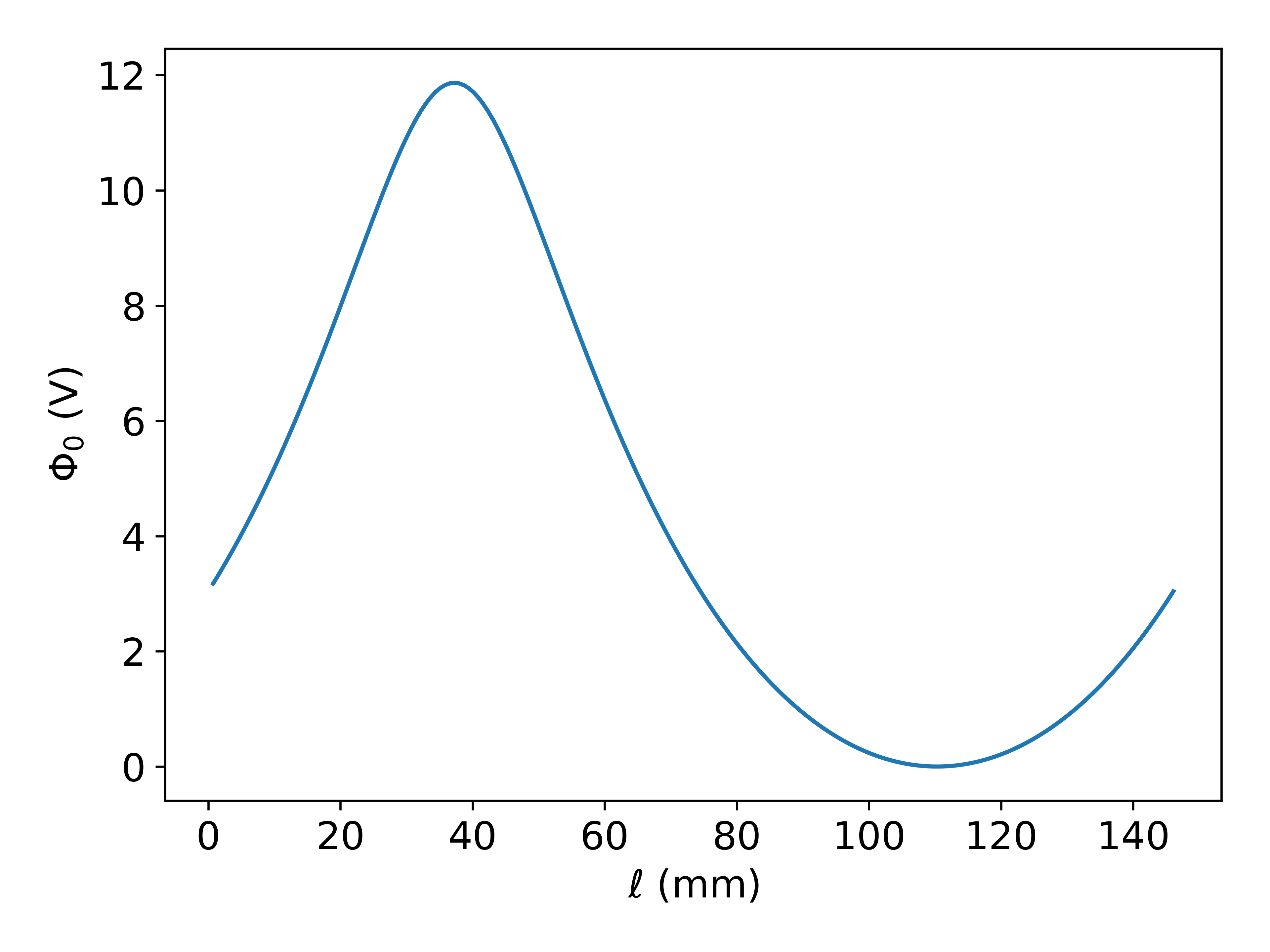}
		\caption{}
		\label{fig:potentialOfEll}
	\end{subfigure}%
	~ \hspace{0mm}
	\begin{subfigure}[t]{0.45\linewidth}
		\centering
		\includegraphics[width = 0.8\linewidth]{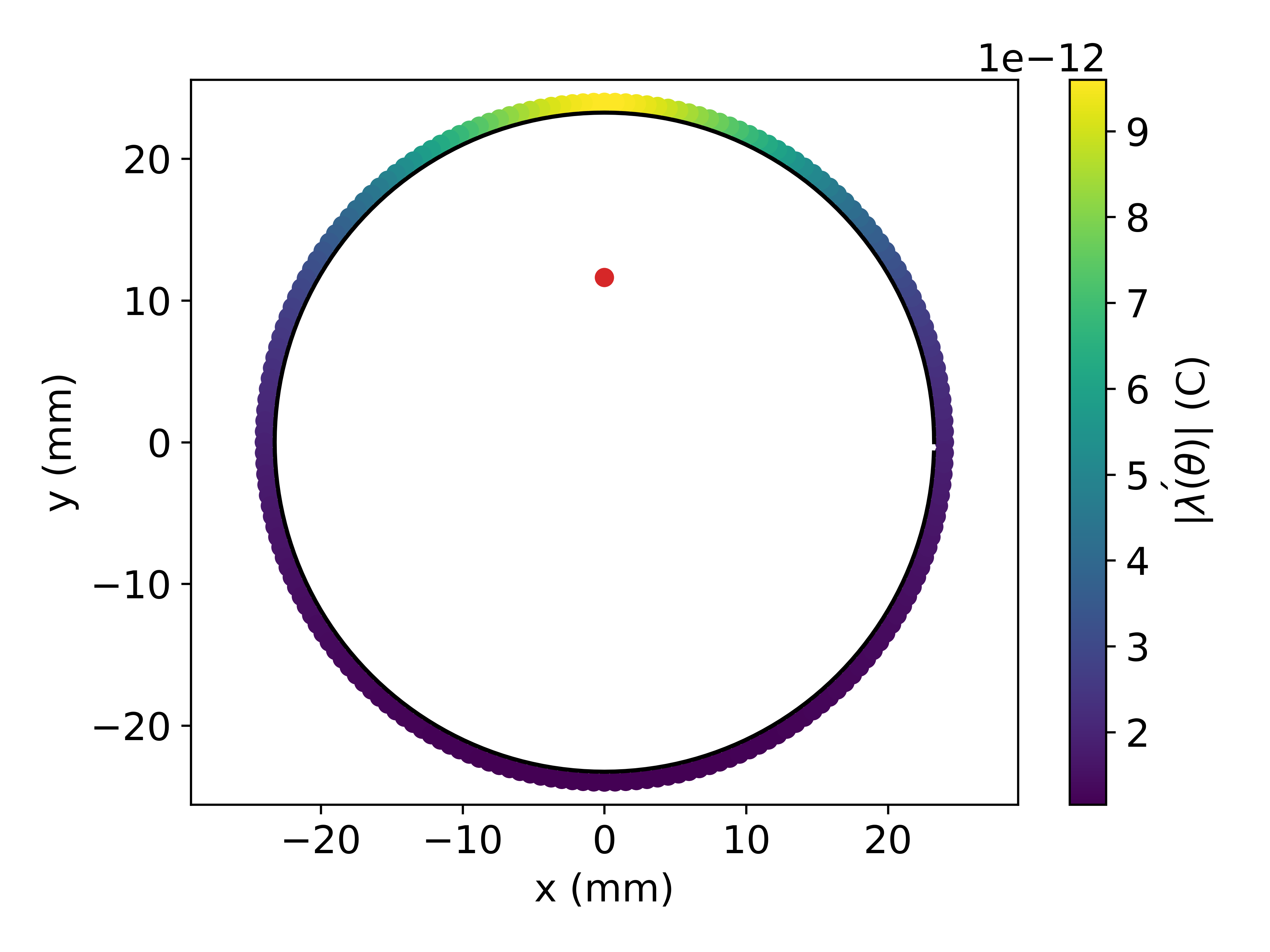}
		\caption{}
		\label{fig:}
	\end{subfigure}
	\caption{\small{Visualization of the method (a) Evolution of the initial potential along the conductor (b) Resulting effective surface charge distribution.}}
	\label{fig:}
\end{figure*}

Our goal is to find an effective surface charge distribution that would give an equal but opposite potential along $\ell$ such that the total potential is zero. If we ground the conductor, electrons will flow from the earth into the conductor because of the positive charge of the beam. Seeing the figure above as an inverted potential well due to the negative charge of electrons, it seems reasonable to state that the linear density of the effective surface charge distribution along the conductor, $\Lambda'(\ell)$ (where $\Lambda'(\ell)d\ell$ has units of C/m, like $\lambda$), should be distributed following the shape of the potential : $ \Lambda' \propto \Phi_0(\ell)$. To obtain a decent order of magnitude, we can furthermore normalize the linear density to the total charge of the beam $\lambda_Q$, and we end up with:
\begin{equation}
\Lambda' (\ell)= -A\cdot \lambda_Q\frac{\Phi_0 (\ell)}{\int \Phi_0 d\ell}
\end{equation}

where $A$ is a proportionality constant. Note that $\Phi_0(\ell)$ could be referenced such that its minimum is zero. In truth, any means to reach a constant potential over the beam screen is a valid approach, provided we respect the considerations of the image charge method.

\subsubsection*{Implementation}

Up to now, we didn't talk about where to place the charge distribution and how to choose the value of $A$. Let's introduce $k =\vec{ r'}/\vec r_B$ to quantify how far from the boundary ($\vec r_B$) we place the effective charge distribution  ($\vec {r'}$). In an ideal case, the charge distribution would be placed on the boundary and so $k\to 1$. For the case of a filament inside a cylindrical boundary, we can note the initial potential as a function of $\theta$ (with $\ell = R \theta$) as :
$$\Phi_0(\theta) = f(\theta) + C_0$$

The potential coming from the surface charge distribution, $\phi$, must therefore be :
$$\phi(\theta) = -f(\theta) + C_1$$

such that $\Phi_0 + \phi$ is constant over the boundary. Setting $C_0 = 0$ for convenience and using the effective surface charge distribution $\Lambda' (\theta)= -\frac{A\cdot \lambda_Q}{\int \Phi_0 d\ell} f(\theta)$, we have :
\begin{align*}
\phi(\theta) &= -\int\frac{\Lambda'(\tau)}{2\pi\varepsilon_0}\ln(\sqrt{(R\cos\theta - kR\cos\tau)^2+(R\sin\theta - kR\sin\tau)^2})(Rd\tau)\\
&=-\frac{R}{2\pi\varepsilon_0 }\int_0^{2\pi}\Lambda'(\tau)\left[\frac{1}{2}\ln(k^2-2k\cos(\theta-\tau)+1) + \ln(R)\right]d\tau
\end{align*}

were $\tau$ is an integration variable. We see that the argument of the logarithm leads to a singularity when $\theta\to\tau$ and $k\to 1$. In fact, setting $k \approx 1$, one can verify that $\frac{1}{2}\ln(k^2-2k\cos(\theta-\tau)+1) \approx -2\pi\ln(2)\delta(\theta-\tau) + \ln(2)$. Under those assumptions, we have:
\begin{align}
\phi(\theta) &\approx \frac{R}{2\pi\varepsilon_0 }\int_0^{2\pi}\Lambda'(\tau)\left[2\pi\ln(2)\delta(\theta-\tau)+ \ln(2R)\right]d\tau\ \nonumber\\
&= \frac{R\ln(2)}{\varepsilon_0 }\Lambda'(\theta) + \int_0^{2\pi}\Lambda'(\tau)\ln(2R)d\tau \nonumber\\
&= -\frac{A\cdot \lambda_Q \ln(2)}{\varepsilon_0\int \Phi_0 d\theta} f (\theta) + C_2
\end{align}

Which leads us to conclude that \textbf{if the charge distribution is indeed proportional to the initial potential, the total potential will be constant over the boundary provided that $\boldsymbol{k\to 1}$ and that}:
\begin{equation}
A \approx \frac{\varepsilon_0\int \Phi_0 d\theta}{\lambda_Q \ln(2)}
\end{equation}

For a general boundary, the potential does not vary solely as a function of $\theta$, but rather as a function of $\ell$. Moreover, since the charge distribution is not necessarily proportional to the initial potential, the precise value of $A$ found above does not provide an absolutely general solution. However, after testing with a few geometries, it seems that the total potential $\Phi_0 + \phi$ obtained via this method is, in general, closer to zero than the initial potential alone. We can therefore think of recursively applying this method to reduce the total potential and eventually converge toward the real solution of our problem. This is why this method is called a \textbf{Relaxation method}. The procedure can be formulated as follow :
\begin{enumerate}
\item Fix $k$ reasonably close to 1 (generally fixed so that the image charges are placed at the same distance from the boundary than the separation between them) and calculate $A$ for the initial potential $\Phi_0$. Both are constant for the rest of the problem

\item Calculate the charge distribution $\Lambda'_i$ and the resulting contribution to the potential $\phi_i$
\item The total potential $\Phi_0 + \phi_i < \Phi_0$ can be seen as the result of a charge $\lambda_{Qi} < \lambda_Q$ placed at the same location.  Estimate this $\lambda_{Qi}$ from the ratio of the potentials $\lambda_{Qi} = \lambda_Q\cdot\frac{\Phi_0+\phi_i}{\Phi_0}$, then calculate a new charge distribution $\Lambda'_{i+1}$ from $\Phi_0 + \phi_i$ and $\lambda_{Qi}$ and add it to the original one
\item Repeat 3 and 4 until the solution converges to the desired precision
\end{enumerate}

\begin{figure*}[!h]
	\captionsetup{width=0.9\textwidth}
	\centering
	
	\begin{subfigure}[t]{0.45\linewidth}
	\hspace{-10mm}
		\centering
		\includegraphics[width = 0.9\linewidth]{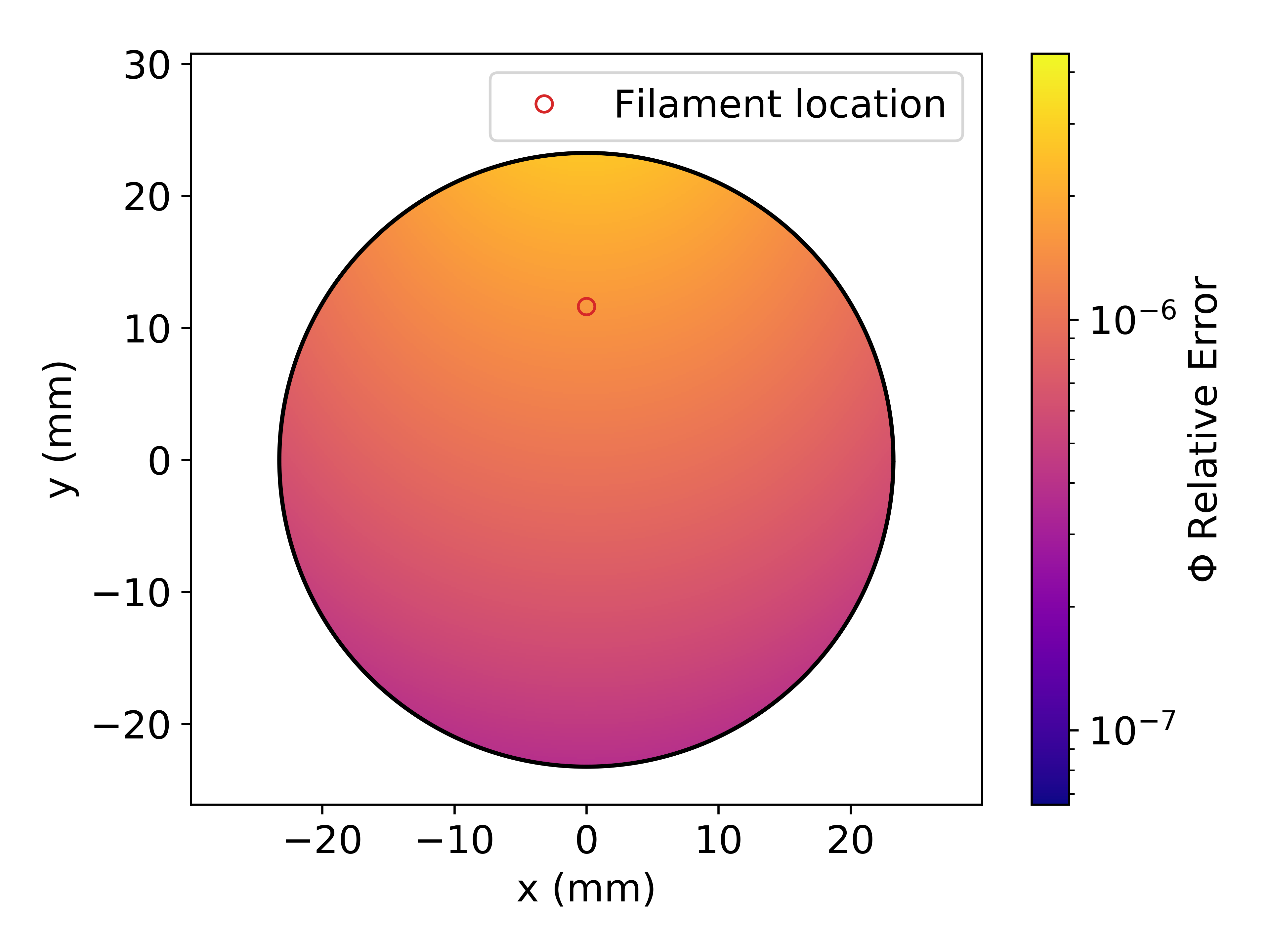}
		\caption{}
		\label{fig:}
	\end{subfigure}%
	~ \hspace{0mm}
	\begin{subfigure}[t]{0.45\linewidth}
		\centering
		\includegraphics[width = 0.9\linewidth]{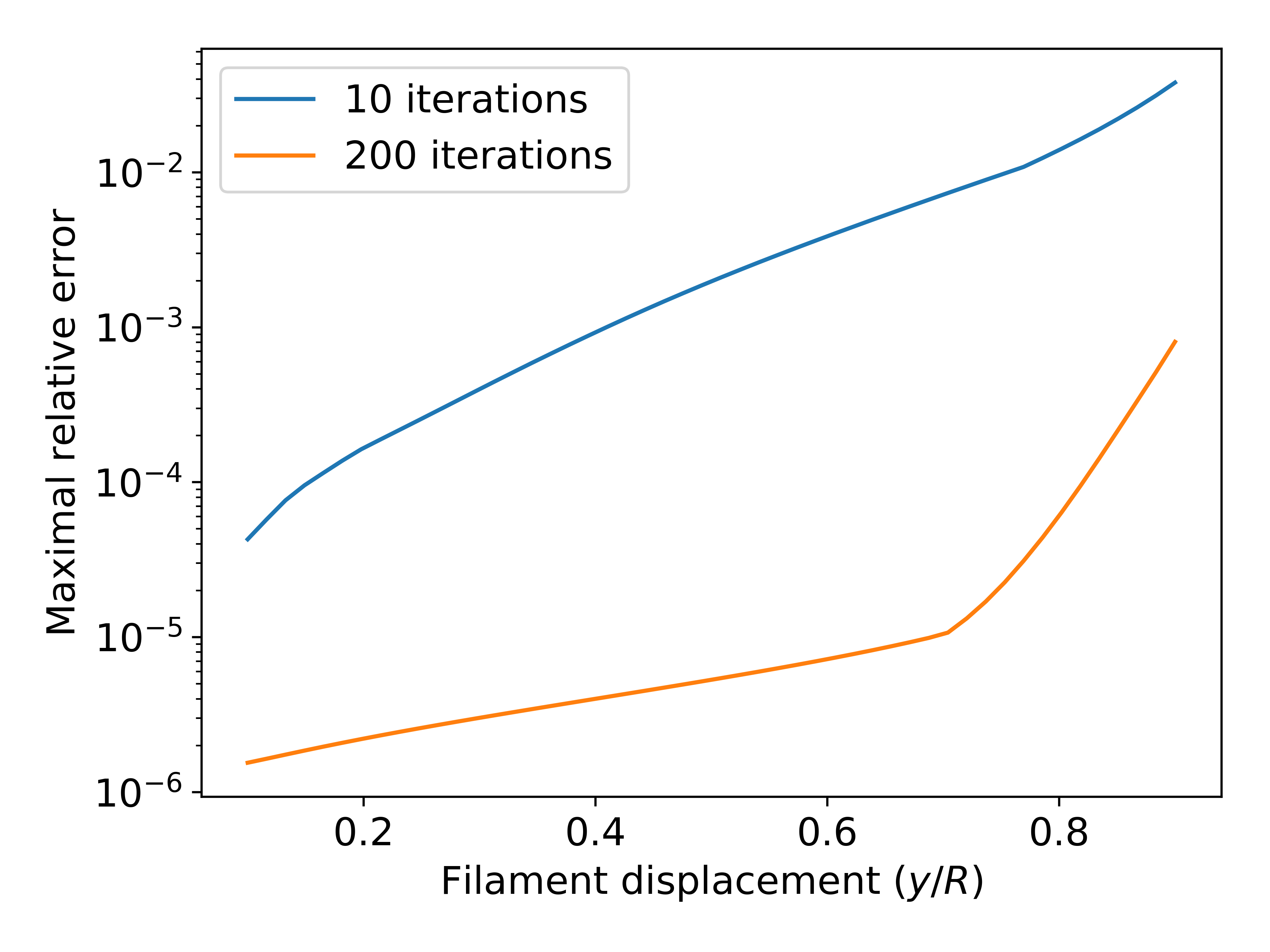}
		\caption{}
		\label{fig:}
	\end{subfigure}
	\caption{\small{Comparison of the method with the analytical solution for a cylindrical conducting boundary. (a) Relative error on the potential over the entire domain for a filament placed midway between the boundary and the center ($y/R = 0.5$) (b) Evolution of the maximal error over the entire domain as a function of the filament displacement from the center. Changing the value of $A$ can make the method converge faster.}\vspace{-10mm}}
	\label{fig:analyticalComp}
\end{figure*}

The method was tested with the problem of a filament inside a cylindrical conduction boundary, for which we know the analytical solution. Figure \ref{fig:analyticalComp} shows an example of the relative error obtained for the potential over the whole physical domain and the evolution of the maximal error when the filament is off-centered. 

%

\newpage
\subsection{Method 2 : Matrix (solving a linear system)}


Let's place, once again, a charge distribution at a position $k\vec r_B$ ($k =\vec{ r'}/\vec r_B$) outside of our physical boundary to compensate for the potential of a real filament $\lambda_Q$ placed inside the boundary. If we place $N$ unknown charges $\lambda_i$, then we can write down $N$ equations specifying the desired potential on the boundary right under a given charge $n$. The $n^\text{th}$ equation is :
\begin{equation}
-\sum_{i=1}^N \frac{\lambda_i}{2\pi\varepsilon_0}\ln(|\vec r_n /k - \vec r_i|) = \frac{\lambda_Q}{2\pi\varepsilon_0}\ln(|\vec r_n/k - \vec r_Q|)
\end{equation}

Or, putting everything in matrix form :
\begin{equation}
-\frac{1}{2\pi\varepsilon_0}\begin{bmatrix}
\ln(|\vec r_1 /k - \vec r_1|) & \ln(|\vec r_1 /k - \vec r_2|) & \dots\\
\ln(|\vec r_2 /k - \vec r_1|)  & \ln(|\vec r_2 /k - \vec r_2|) & \dots \\
\vdots & \vdots & \ddots
\end{bmatrix} \begin{bmatrix}
\lambda_1 \\ \lambda_2 \\ \vdots
\end{bmatrix} = \frac{\lambda_Q}{2\pi\varepsilon_0}\begin{bmatrix}\ln(|\vec r_1/k - \vec r_Q|)\\ \ln(|\vec r_2/k - \vec r_Q|)
\\\vdots\end{bmatrix}
\end{equation}

where $\vec r_Q$ is the position of the real charge inside the boundary. We want the potential coming from all the charges in the discretized charge distribution to compensate for the potential coming from the real charge.

\subsubsection*{Implementation}
The implementation of this method is as simple as it seems. We only need to generate the $N\times N$ matrix associated with the system of equations given above and solve this linear system in order to find the precise values of all the $\lambda_i$ using any conventional inversion algorithm. In truth, this method gives the \textbf{exact} solution to the problem of a discretized boundary, which happens to be really close to the exact solution of the actual continuous boundary (R. Baartman's words). \textbf{This method should be preferred to the first one, as it will generally give better results and is more robust}. However, the method only works when the $N\times N$ matrix is invertible and will use considerably more memory than the first one for large values of $N$. Below is a comparison of the error made in the case of the conducting cylindrical boundary for the two methods. 

\begin{figure*}[!h]
	\captionsetup{width=0.9\textwidth}
	\centering
	
	\begin{subfigure}[t]{0.45\linewidth}
	\hspace{-10mm}
		\centering
		\includegraphics[width = 0.8\linewidth]{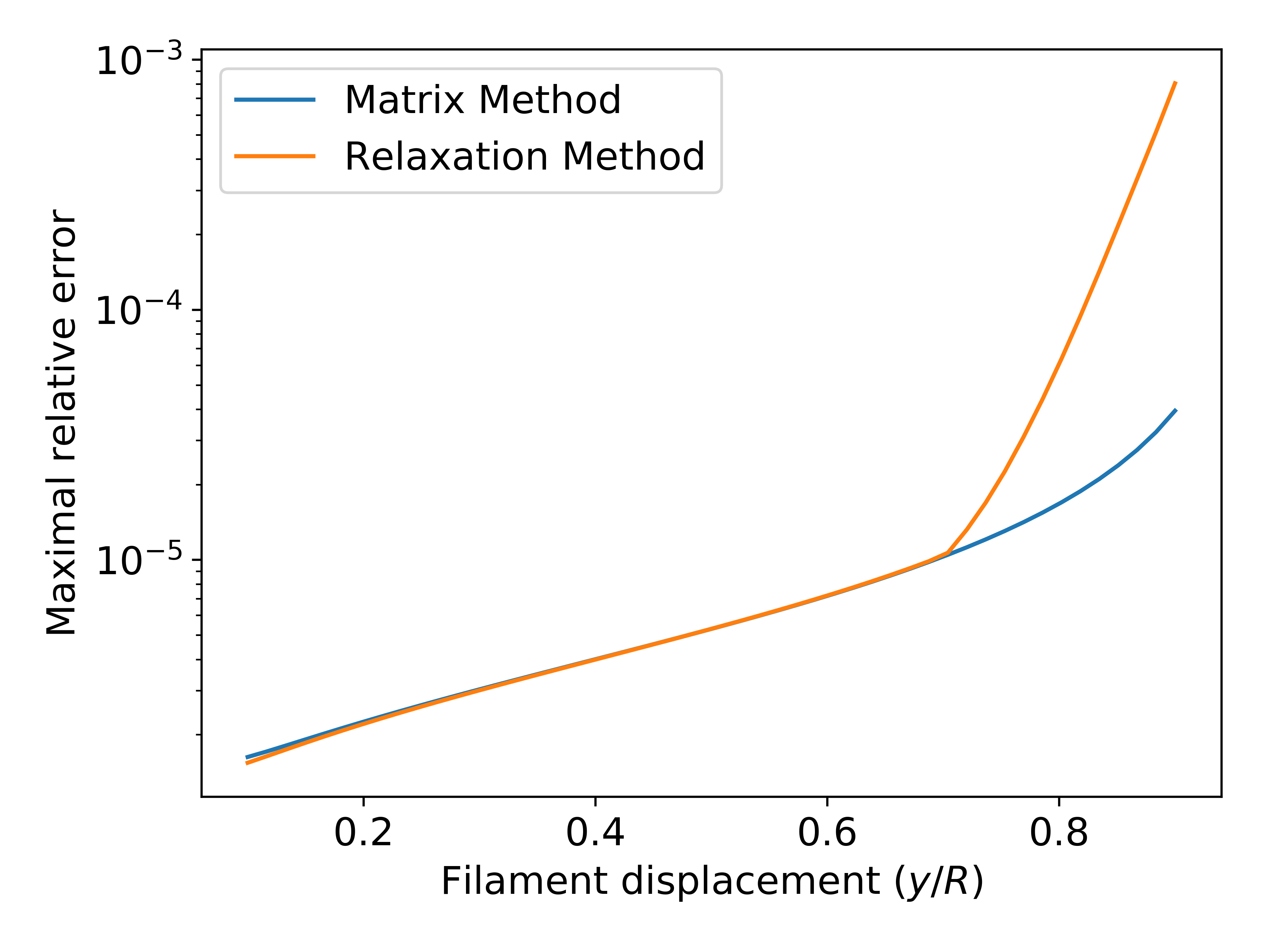}
		\caption{}
		\label{fig:}
	\end{subfigure}%
	~\hspace{0mm}
	\begin{subfigure}[t]{0.45\linewidth}
		\centering
		\includegraphics[width = 0.8\linewidth]{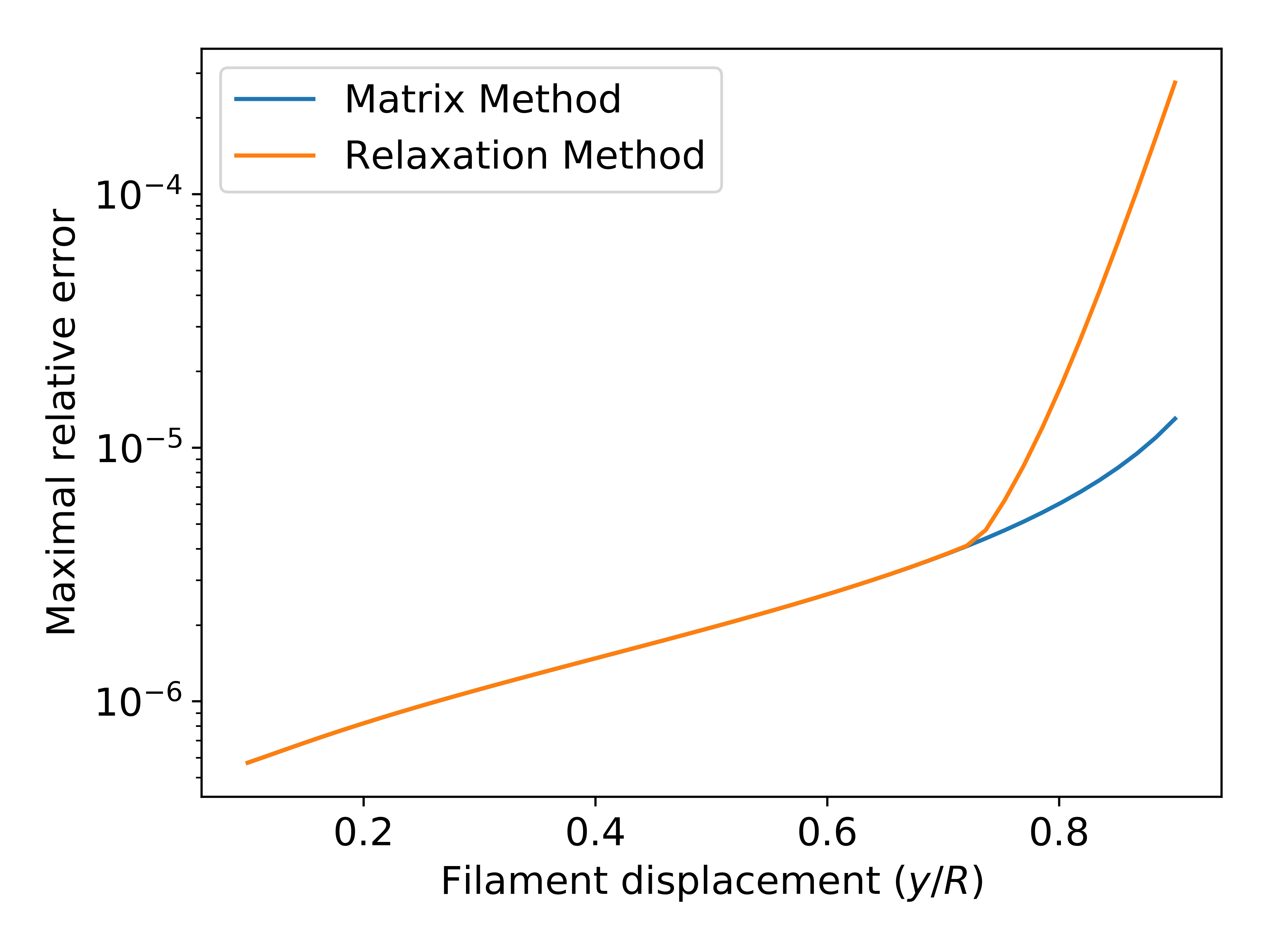}
		\caption{}
		\label{fig:}
	\end{subfigure}
	\caption{\small{Comparison of the two methods. (a) $N=200$ (b) $N=500$. The relaxation method was applied with 200 iterations. It is believed that more iterations (or tweaked $A$ value) would lead to a perfect agreement between the two methods.}\vspace{-10mm}}
	\label{fig:}
\end{figure*}

\newpage
\subsection{The LHC case : centered beam}
With the LHC beam centered in the beam screen, we get the initial potential shown on figure \ref{fig:potentialOfEll}.

\begin{figure*}[!h]
	\captionsetup{width=0.9\textwidth}
	\centering
	
	\begin{subfigure}[t]{0.45\linewidth}
	\hspace{-10mm}
		\centering
		\includegraphics[width = 0.8\linewidth]{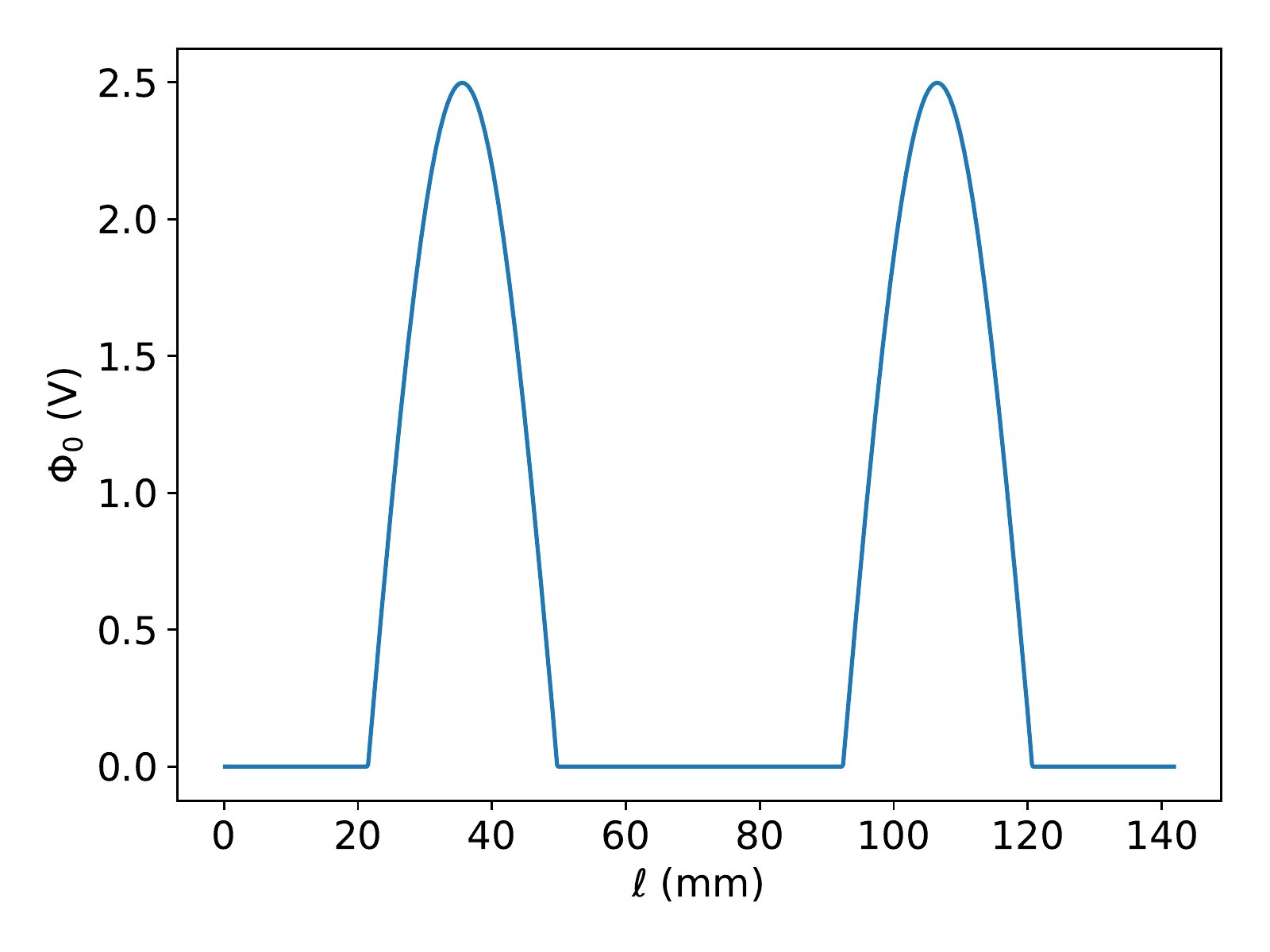}
		\caption{}
		\label{fig:potentialOfEll}
	\end{subfigure}%
	~ \hspace{0mm}
	\begin{subfigure}[t]{0.45\linewidth}
		\centering
		\includegraphics[width = 0.8\linewidth]{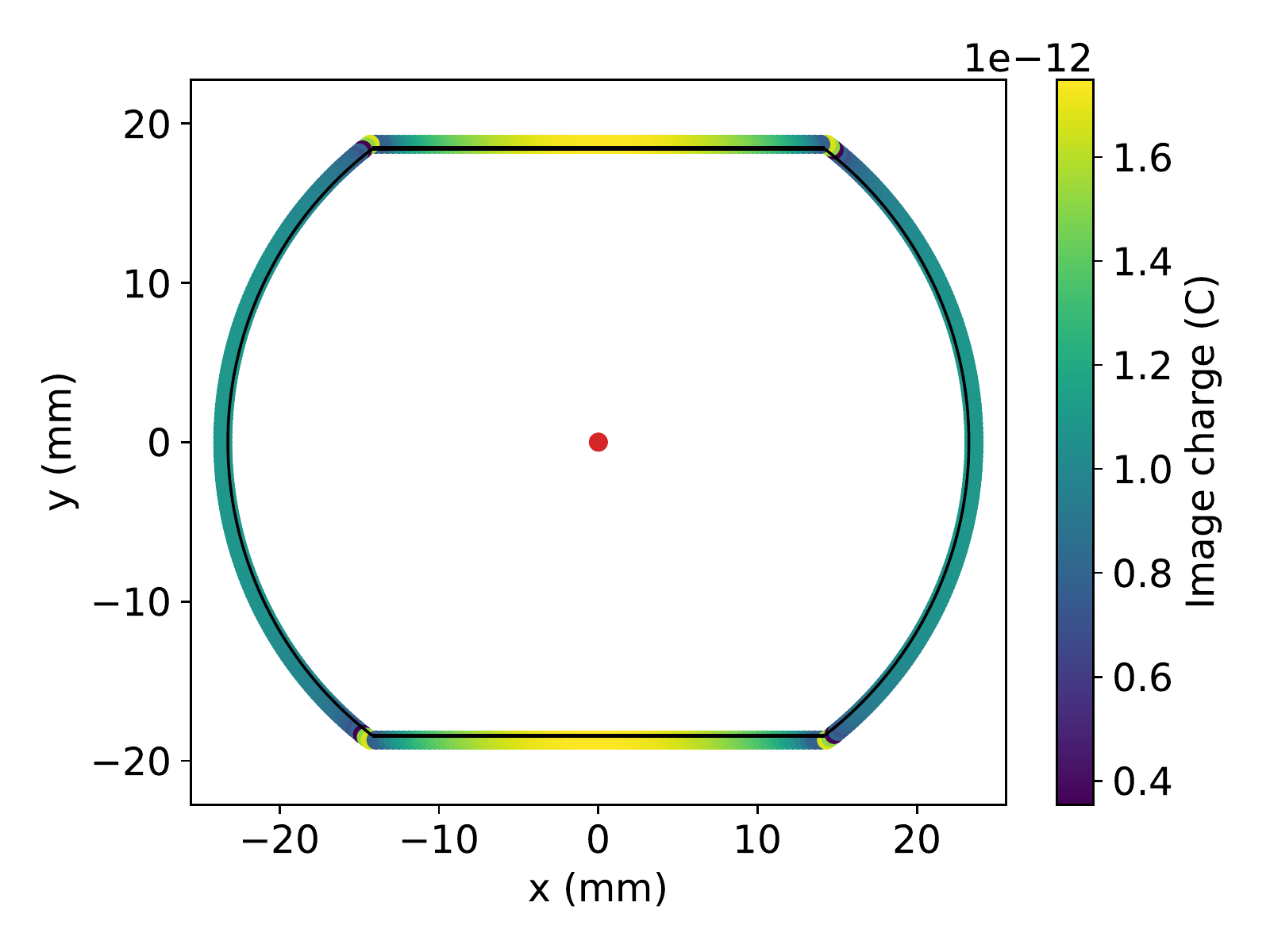}
		\caption{}
		\label{fig:}
	\end{subfigure}
	\caption{\small{Visualisation of the method (a) Evolution of the initial potential along the LHC beam screen (b) Resulting effective surface charge distribution obtained with the matrix method.}}
	\label{fig:}
\end{figure*}

Applying any of the methods presented above, we need either $N=200$ for the matrix method or 200 iterations of the relaxation method to reach a sufficiently flat potential along the beam screen. The figure below shows the equipotential lines around the edge of the beam screen, and a comparison between the initial potential $\Phi_0(\ell)$ along the beam screen and the total potential.

\begin{figure*}[!h]
	\captionsetup{width=0.9\textwidth}
	\centering
	
	\begin{subfigure}[t]{0.45\linewidth}
	\hspace{-10mm}
		\centering
		\includegraphics[width = 0.9\linewidth]{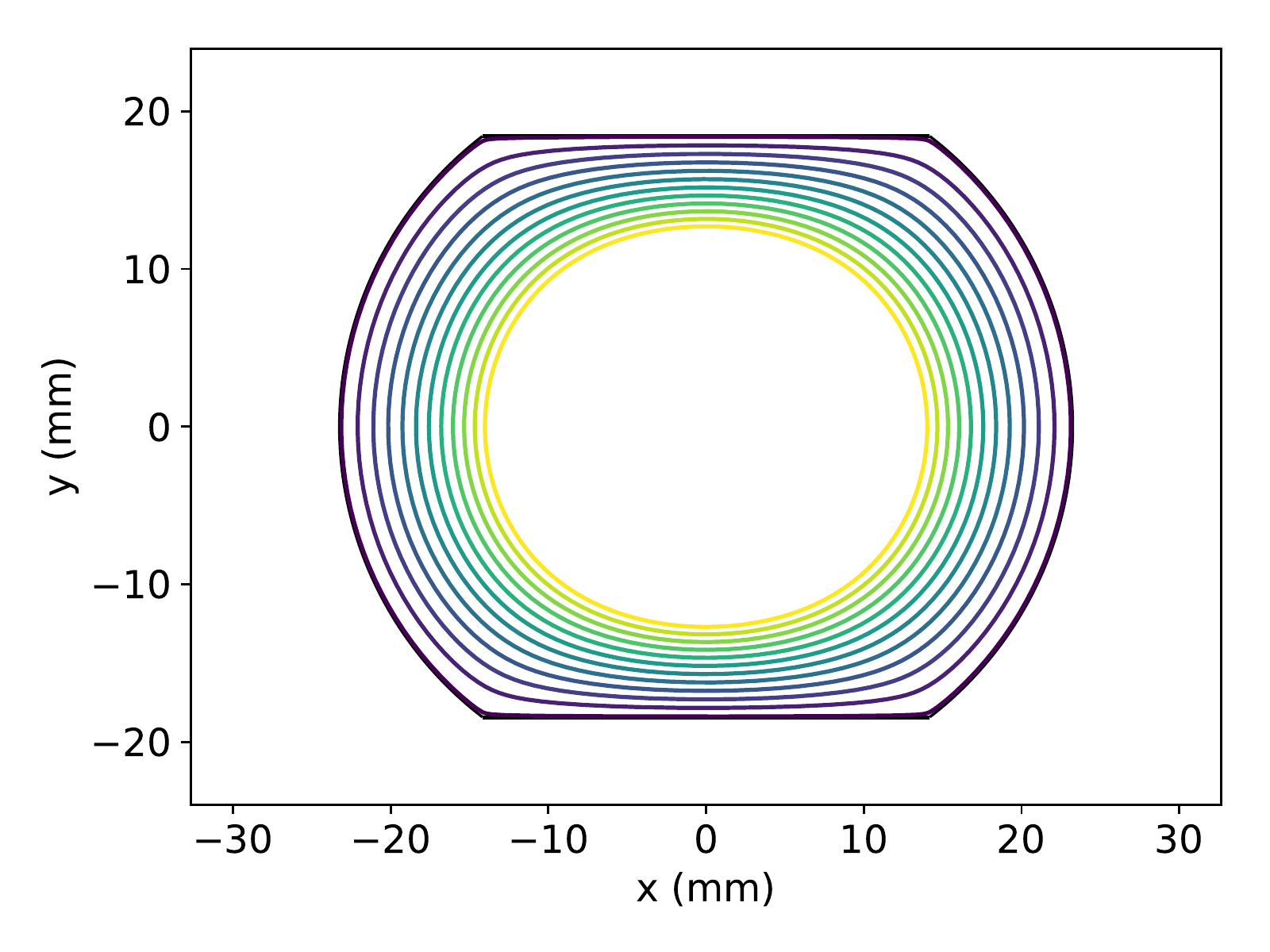}
		\caption{}
		\label{fig:}
	\end{subfigure}%
	~\hspace{0mm}
	\begin{subfigure}[t]{0.45\linewidth}
		\centering
		\includegraphics[width = 0.9\linewidth]{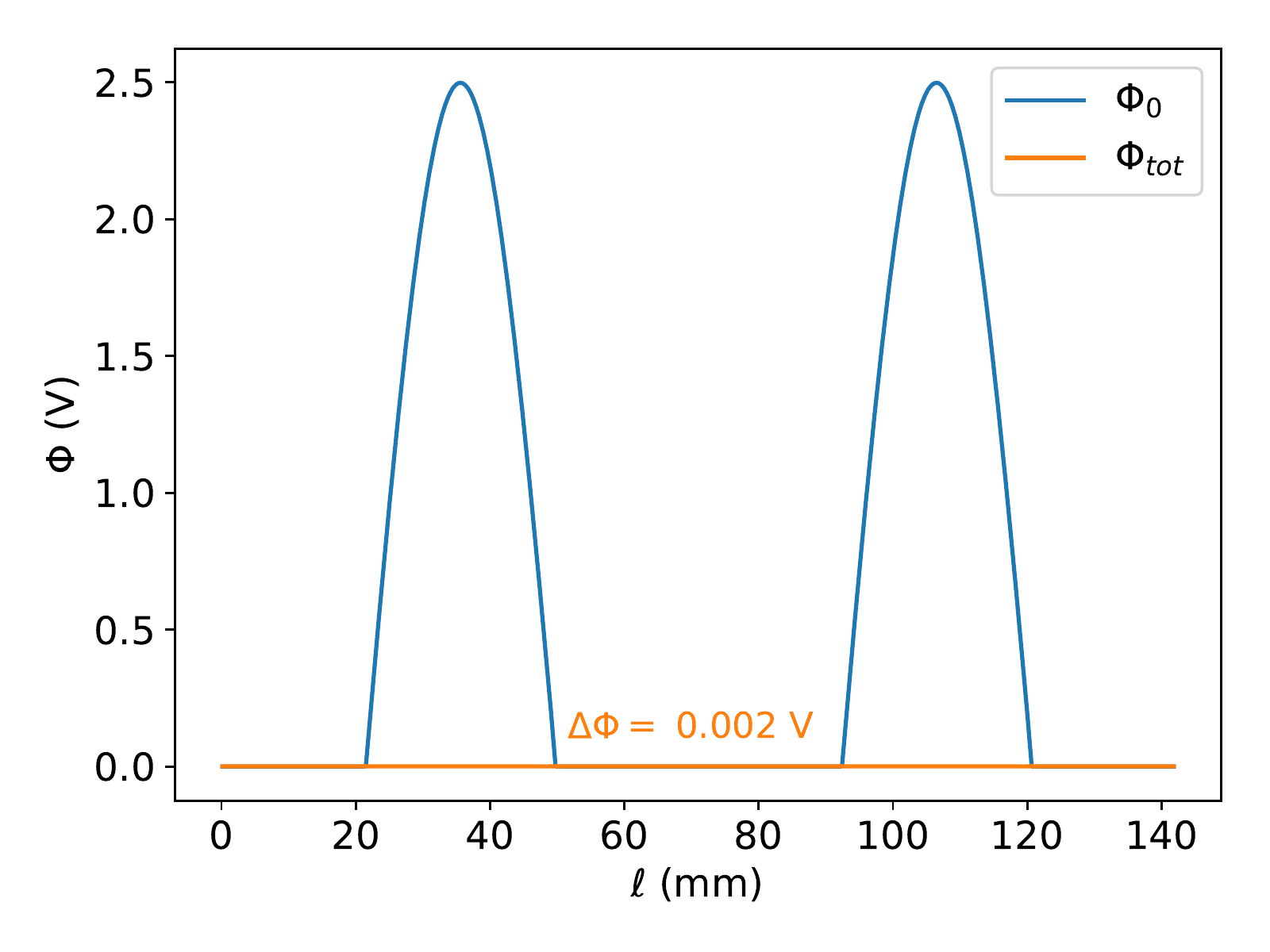}
		\caption{}
		\label{fig:}
	\end{subfigure}
	\caption{\small{Potential obtained with the matrix method for the LHC beam in the LHC beam screen. (a) Equipotential lines around the beam screen  (b) Comparison between the free space potential along the beam screen and the total potential obtained.}}
	\label{fig:}
\end{figure*}

\begin{figure*}[!h]
	\captionsetup{width=0.9\textwidth}
	\centering
	\begin{subfigure}[t]{0.45\linewidth}
		\centering
		\includegraphics[width = 1\linewidth]{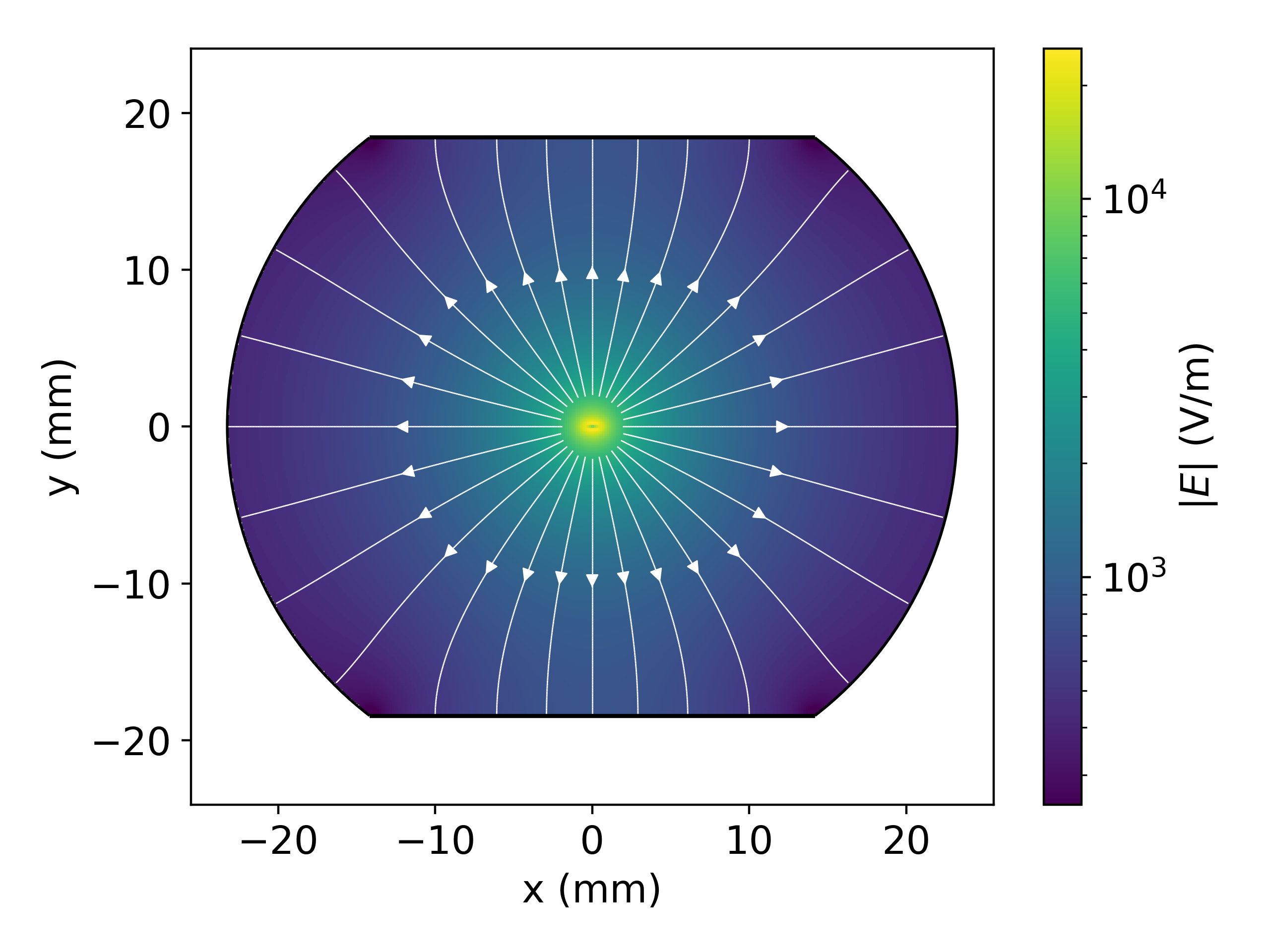}
		\caption{}
		\label{fig:totalField}
	\end{subfigure}%
	~\hspace{7mm}
	\begin{subfigure}[t]{0.45\linewidth}
		\centering
		\includegraphics[width = 1\linewidth]{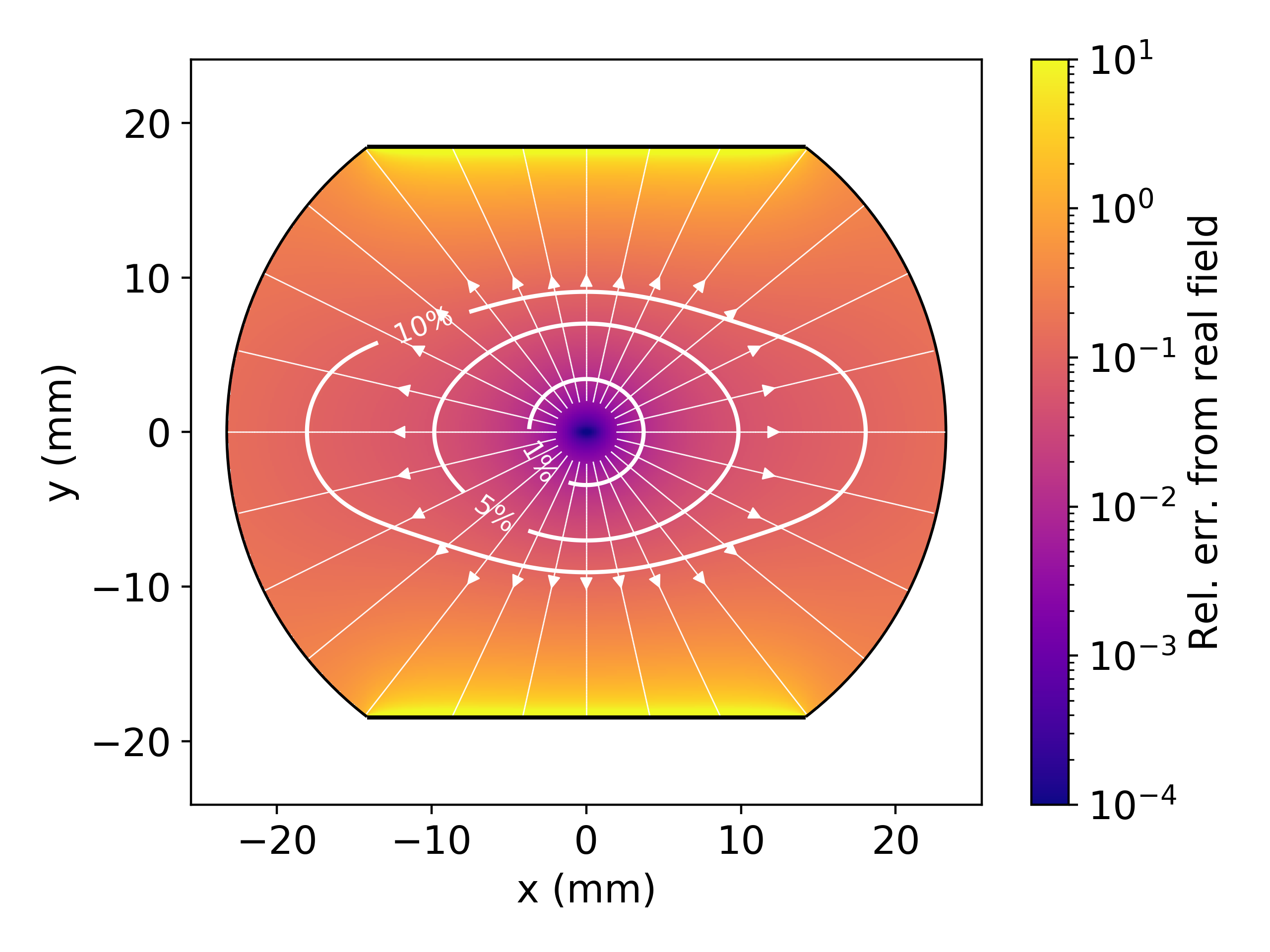}
		\caption{}
		\label{fig:RESULT}
	\end{subfigure}
	\caption{\small{Comparison of (a) the total electric field around the beam in presence of the beam screen (b) the free space Houssais E-field. The magnitude of the relative error vector (free space field compared with the total field) is shown on the color bar of (b). The level curves for errors of 1\%, 5\% and 10\% are shown. Outside of those regions, the free space electric field differs greatly from the total electric field and is no longer a good approximation.}}
	\label{fig:finalField}
\end{figure*}

As seen above, the results are much better than the ones obtained with the image charge method presented in the previous section. We can again compute the electric field obtained and compare it with the free space approximation. The results are shown on figure \ref{fig:finalField}.

\newpage
\subsection{The LHC case : off-centered beam}

Both methods presented in this note apply equally well to the case of an off-centered beam, even with the complex geometry of the LHC beam pipe. Below are presented the potential, as well as the electric field and the effective surface charge distribution for the off-centered beam. All those results were obtained with the matrix method with $N=500$.

\begin{figure*}[!h]
	\captionsetup{width=0.9\textwidth}
	\centering
	
	\begin{subfigure}[t]{0.45\linewidth}
	\hspace{-10mm}
		\centering
		\includegraphics[width = 0.9\linewidth]{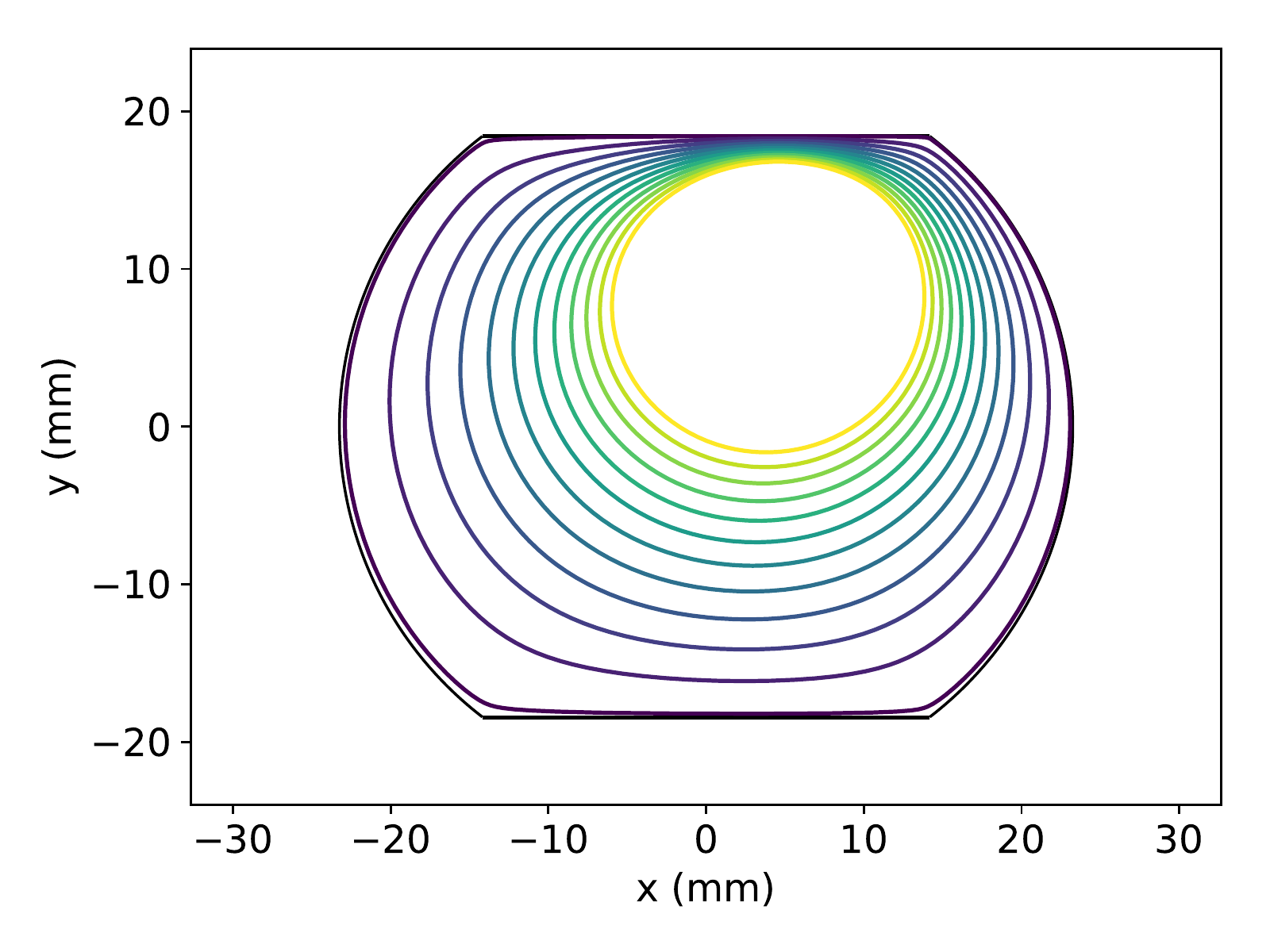}
		\caption{}
		\label{fig:}
	\end{subfigure}%
	~\hspace{0mm}
	\begin{subfigure}[t]{0.45\linewidth}
		\centering
		\includegraphics[width = 0.9\linewidth]{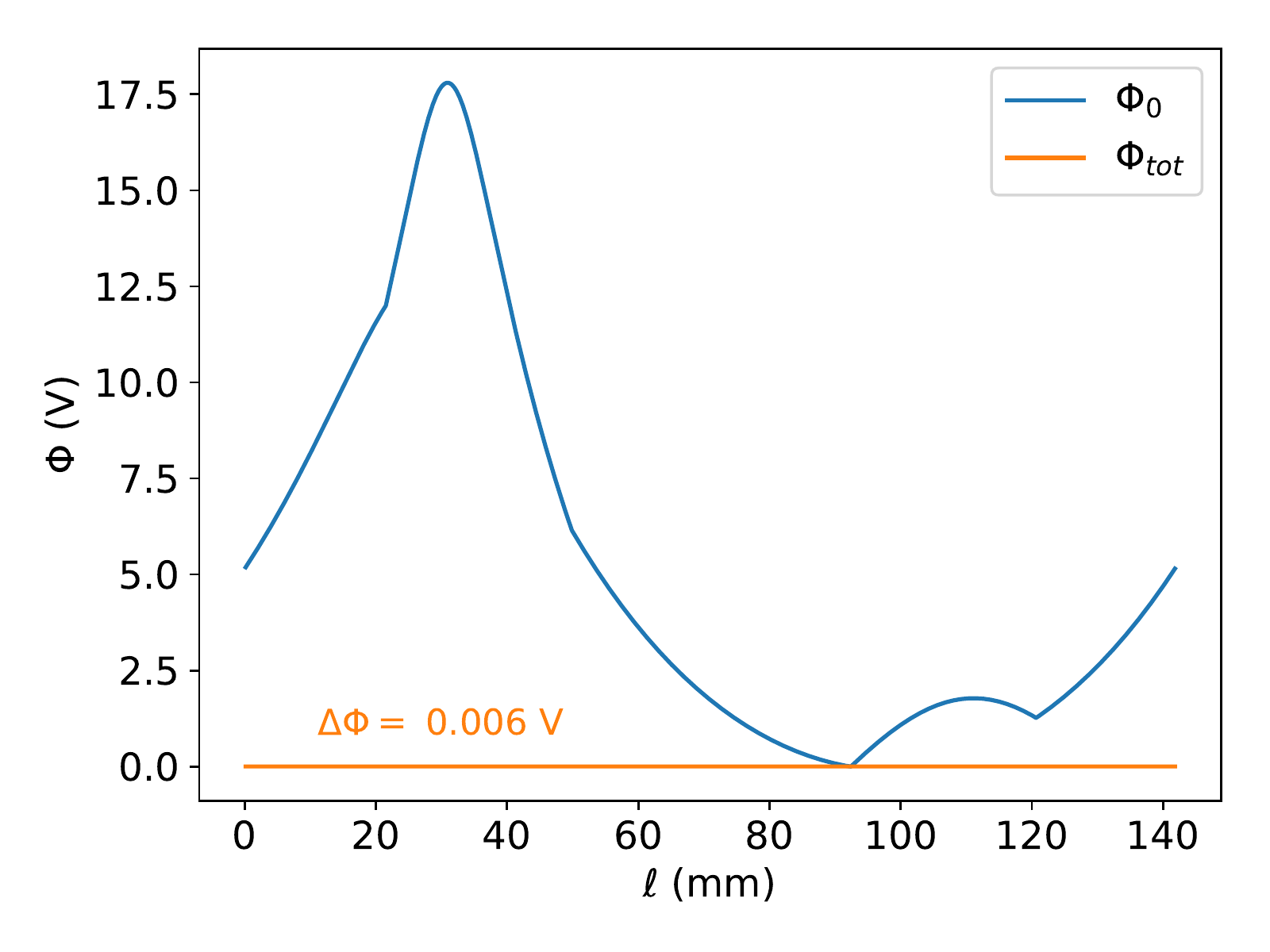}
		\caption{}
		\label{fig:}
	\end{subfigure}
	\caption{\small{Potential obtained with the matrix method for the off-centered LHC beam in the LHC beam screen. (a) Equipotential lines around the beam screen  (b) Comparison between the free space potential along the beam screen and the total potential obtained.}}
	\label{fig:}
\end{figure*}

\begin{figure*}[!h]
	\captionsetup{width=0.9\textwidth}
	\centering
	\begin{subfigure}[t]{0.45\linewidth}
		\centering
		\includegraphics[width = 1\linewidth]{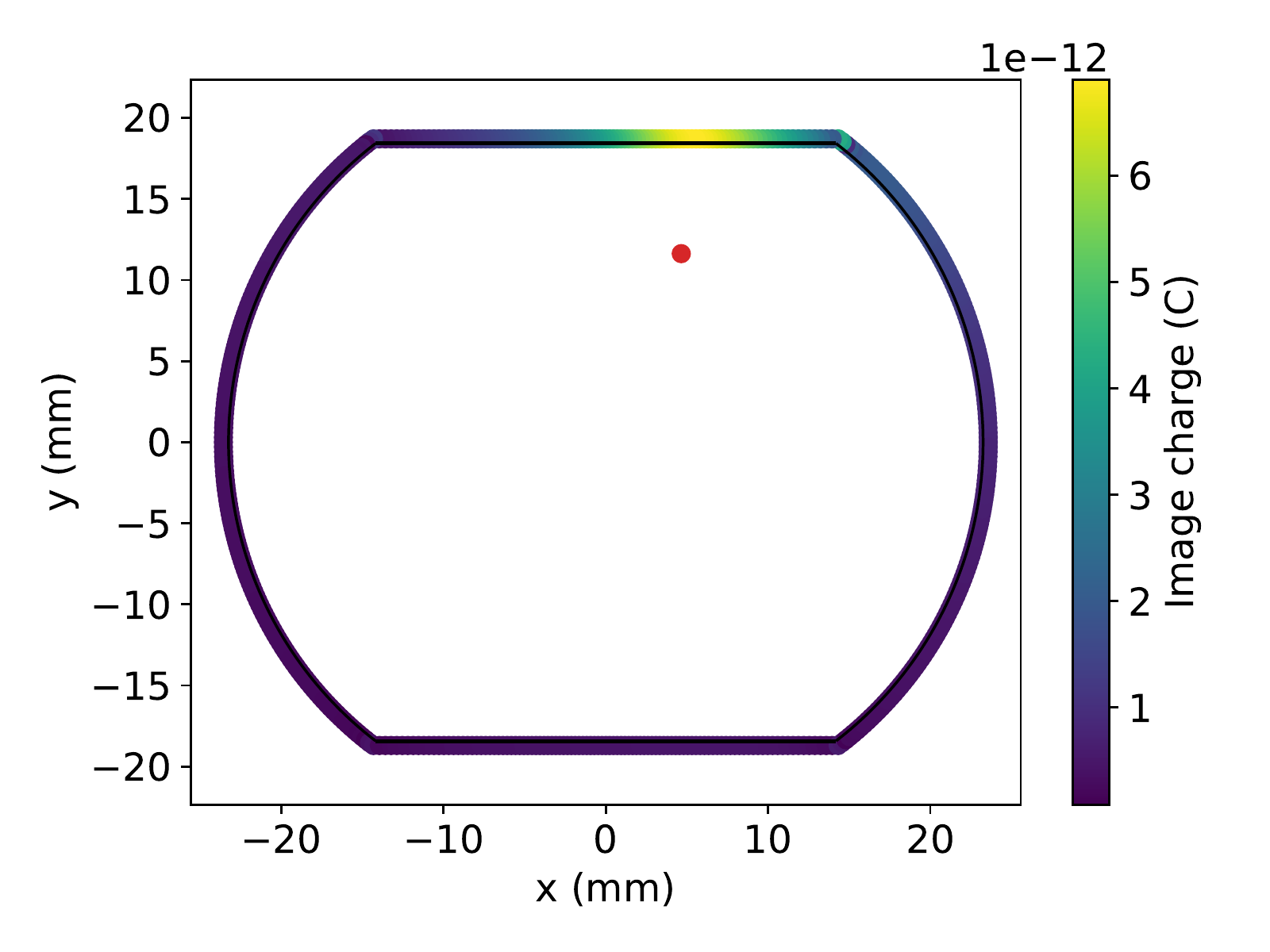}
		\caption{}
		\label{fig:}
	\end{subfigure}%
	~\hspace{7mm}
	\begin{subfigure}[t]{0.45\linewidth}
		\centering
		\includegraphics[width = 1\linewidth]{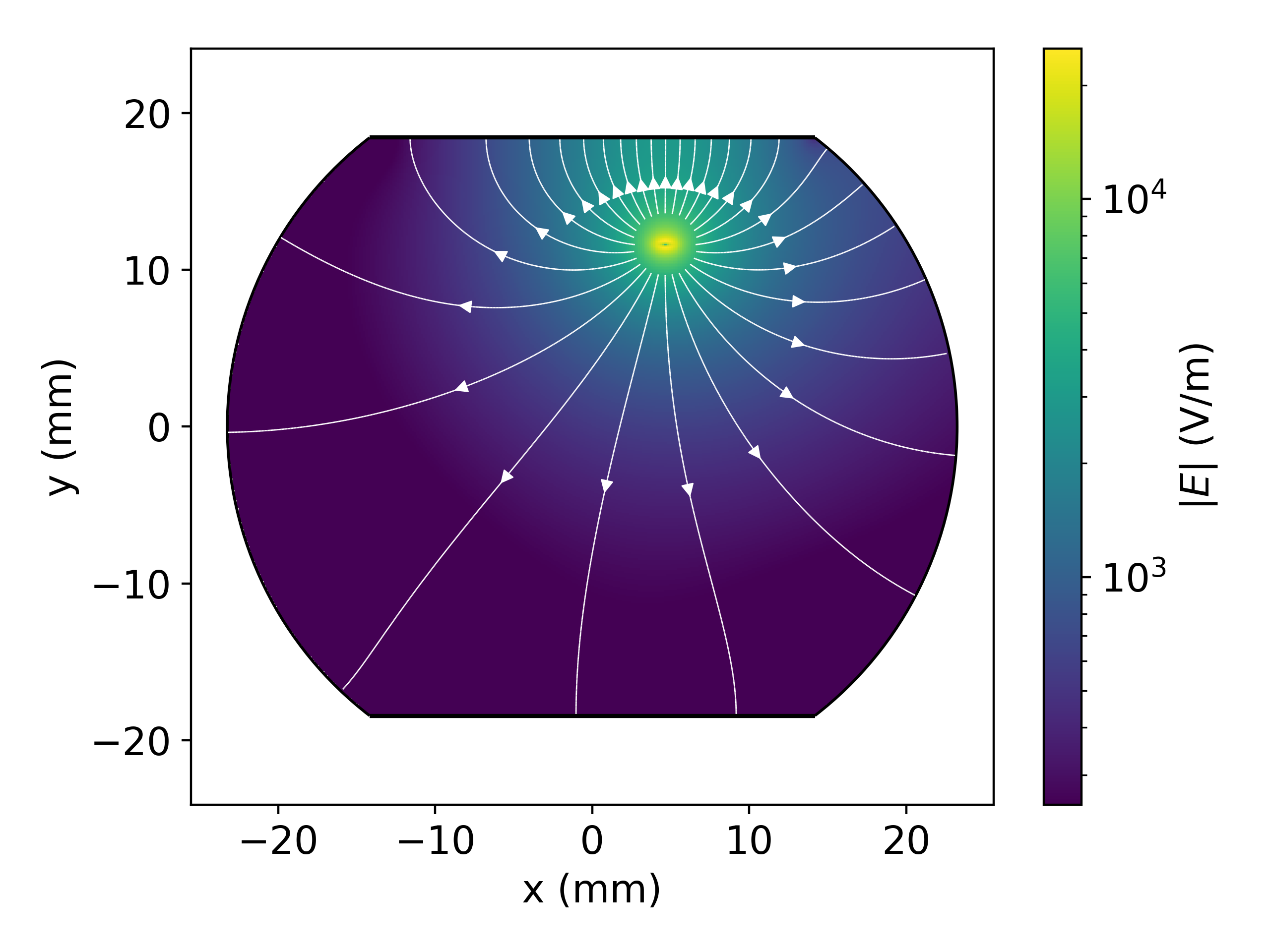}
		\caption{}
		\label{fig:}
	\end{subfigure}
	\caption{\small{(a) Effective surface charge distribution for the off-centered LHC beam (b) Total electric field around the beam in presence of the beam screen. The error coming from only using the free space field in this case is beyond 100\% on almost all the domain.}}
	\label{fig:}
\end{figure*}

\newpage
\section{Conclusion}
The above calculations were intended to show the importance of considering the LHC beam screen when computing the E-field around the LHC beam. Using the method of images, we found a charge distribution which is able to mimic (to some degree) the contribution of the beam screen on the E-field surrounding the LHC beam. We then developed two new methods for finding the effect of a conducting boundary in electrostatics boundary value problems. We have shown that those two methods are robust, yielding good results no matter where the real charge is placed inside the boundary. Ultimately, we have shown that in the case of the LHC beam, using the free space E-field of a 2D Gaussian beam is only accurate to 1\% for locations closer than $10\sigma$ from the center of the beam, and only accurate to 10\% for locations closer than $30\sigma$.

\addcontentsline{toc}{section}{\protect\numberline{References}}

 \nocite{Schaich1991}
\bibliographystyle{IEEEtran}
\bibliography{references}
\end{document}